\documentclass[a4paper,11pt]{article}
\usepackage{jheppub}
\usepackage{float}
\usepackage{color}
\usepackage{mymacros}
\usepackage{chngcntr}
\usepackage{subcaption}
\usepackage{verbatim}
\usepackage{amsthm}
\usepackage{graphics}
\usepackage{graphicx}
\usepackage{xcolor}
\usepackage{amsmath}
\usepackage{amssymb}
\usepackage{mathrsfs}
\usepackage{bbold}
\usepackage{cases}
\usepackage{epsfig}
\usepackage{epstopdf}
\usepackage{hyperref}
\usepackage{booktabs}

\title{\boldmath Complexity equals anything for multi-horizon black holes}

\author{Hong-Yue Jiang $^{a\,b\,c}$ \footnote{jianghy21@lzu.edu.cn}}
\author{and Yu-Xiao Liu$^{a\,b\,c}$ \footnote{liuyx@lzu.edu.cn, corresponding author}}

\affiliation{
$^{a}$Lanzhou Center for Theoretical Physics, Key Laboratory of Theoretical Physics of Gansu Province, Key Laboratory for Quantum Theory and Applications of the Ministry of Education, Lanzhou University, Lanzhou, Gansu 730000, China\\
$^{b}$Institute of Theoretical Physics $\&$ Research Center of Gravitation, School of Physics Science and Technology, Lanzhou University, Lanzhou 730000, China\\
$^{c}$Gansu Provincial Research Center for Basic Disciplines of Quantum Physics, Lanzhou University, Lanzhou, Gansu 730000, China
}

\abstract{We investigate the ``complexity equals anything" proposal with codimension-one and codimension-zero gravitational observables for multi-horizon black holes, using the Bardeen-AdS class black hole as an example. 
In particular, we compare the results with the ``complexity equals volume" (CV) proposal and find that the 
 {generalized volume complexitiy} enables the probing of a more complete black hole interior, that is, all spacetime regions where the blackening factor $f(r)<0$. 
This is the advantage brought by the flexibility of this holographic complexity conjecture. 
In addition, we compute the 
 {codimension-zero gravitational observables} derived from various geometric quantities and show that these constructions can effectively differentiate the distinct interior regions of the black hole. 
}

\begin{document}
\maketitle
\section{Introduction}\label{sec1}

In recent years, the holographic principle has attracted widespread attention as a good perspective for understanding the connections between quantum gravity and quantum field theory. 
The AdS/CFT correspondence~\cite{Maldacena:1997re,Witten:1998qj} allows us to quantitatively analyze the relation between the geometric quantities in the bulk spacetime and the entanglement properties of the boundary field theory~\cite{Ryu:2006bv,Ryu:2006ef,Maldacena:2013xja}. 
Quantum circuit complexity measures the difficulty of preparing boundary quantum states from a simple reference state~\cite{Jefferson:2017sdb,Chapman:2017rqy,Bernamonti:2019zyy,Brown:2017jil}. 
The definition of circuit complexity is ambiguous, because of the flexibility in the reference state or the quantum gates. 
In terms of the holography, quantum circuit complexity is considered to be dual to the bulk gravitational theory~\cite{Susskind:2014rva,Susskind:2014moa}. 
Recently, Krylov complexity has emerged as another candidate for describing holographic complexity~\cite{Rabinovici:2023yex, Nandy:2024htc}. 
A review paper published this year provides a comprehensive overview of recent progress in quantum complexity and holographic complexity~\cite{Baiguera:2025dkc}.

Holographic complexity is widely regarded as an excellent probe of the interior of black holes. 
It is most commonly explored in the context of the thermofield double (TFD) state, which is dual to an eternal two-sided black hole~\cite{Maldacena:2001kr}. 
There are three important proposals regarding holographic complexity: 
complexity equals volume (CV)~\cite{Susskind:2014rva}, complexity equals action (CA)~\cite{Brown:2015bva}, and complexity equals spacetime volume (CV 2.0)~\cite{Couch:2016exn}. 
Each of these proposals satisfies the basic characteristics of quantum circuit complexity. 
First, complexity grows linearly at late times~\cite{Susskind:2014moa}. 
Second, the response of complexity to perturbations exhibits a universal time delay, commonly referred to as the switchback effect~\cite{Stanford:2014jda}. 
In addition, all of the holographic complexity proposals have ambiguous definitions, such as the ill-defined length scale $l$ in CV and CV 2.0, or a similarly indeterminate length scale that occurs on the null boundaries of the Wheeler-de Witt patch. 
It was pointed out in Refs.~\cite{Belin:2021bga,Belin:2022xmt,Jorstad:2023kmq,Myers:2024vve} 
that the ambiguities in quantum circuit complexity can be naturally matched by those in its holographic dual, i.e., the ``complexity equals anything'' proposal ($\mathcal{C}_{\text{gen}}$). 
This proposal suggests that complexity can be dual to a series of gravitational observables. 
Concrete implementations of this proposal can be found in several examples (e.g.~\cite{Caceres:2023ziv,Wang:2023noo,Zhang:2024mxb,Omidi:2022whq,Wang:2023eep,Jiang:2023jti,Arean:2024pzo,Emami:2024mes,Caceres:2025myu,Aguilar-Gutierrez:2023zqm,Aguilar-Gutierrez:2024rka}). 
In recent years, there has been significant progress toward 
holographic complexity~\cite{Cai:2016xho,Goto:2018iay,Jorstad:2022mls,Jiang:2018pfk,Hernandez:2020nem,Yang:2017hbf,Miyaji:2025yvm,Guo:2017rul,Chapman:2018dem,Jiang:2018sqj,Carmi:2017jqz,Zolfi:2023bdp,Lehner:2016vdi,Chang:2024muq,Fu:2024vin,Frey:2024tnn,Pedraza:2022dqi, Carrasco:2023fcj, Aguilar-Gutierrez:2023tic, Caceres:2024edr,Pedraza:2021fgp,Pedraza:2021mkh,Emparan:2021hyr,Chen:2023tpi,Aguilar-Gutierrez:2023ccv,An:2022lvo,Brown:2015lvg,Carmi:2016wjl,Baiguera:2023tpt,Baiguera:2024xju,Miyaji:2025jxy}, 
quantum circuit complexity~\cite{Caceres:2019pgf,Brown:2019whu,Guo:2018kzl,Ruan:2020vze,Bernamonti:2020bcf,Hackl:2018ptj,Qu:2021ius,Qu:2022zwq,Jin:2025fpa,Caputa:2017yrh},  
and Krylov complexity~\cite{Balasubramanian:2022tpr,Caputa:2024vrn,Baggioli:2024wbz,Heller:2024ldz,Rabinovici:2021qqt,Jian:2020qpp,Huh:2023jxt, Huh:2024lcm, Jeong:2024oao, Baggioli:2025ohh, Bhattacharjee:2022vlt, Das:2024zuu, Xu:2024gfm, Aguilar-Gutierrez:2025mxf}.

 {A major advantage of the ``complexity equals anything" proposal is that its construction allows extremal surfaces to approach a spacelike singularity, thereby probing geometric features near the singularity. 
This naturally raises a question: Does the ``complexity equals anything" proposal possess an absolute advantage in uncovering the internal structure of a black hole? 
To address this, we aim to analyze black holes with more complex internal structures: such as the hairy black hole discussed in Ref.~\cite{Arean:2024pzo}, or the multi-horizon black hole that we will discuss in this work. 
On the other hand, }in our previous papers~\cite{Wang:2023eep,Jiang:2023jti}, we considered the ``complexity equals anything" proposal with codimension-one gravitational observables for the Reissner-Nordström-AdS black hole and the charged Bañados-Teitelboim-Zanelli black hole. 
Both of them are static solutions to the Einstein-Maxwell field equations. 
In our opinion, it is necessary to discuss black holes in Einstein gravity coupled to nonlinear electrodynamics, which goes beyond the standard Maxwell framework, such as the Bardeen-AdS class black hole. 
In this paper, we explore the ``complexity equals anything” proposal in the context of the Bardeen-AdS class black hole, employing both codimension-one and codimension-zero gravitational observables. 
Unlike the Maxwell field, the nonlinear electromagnetic field results in a significantly richer and more intricate internal structure, often characterized by the presence of multiple horizons. 
Our analysis demonstrates that the ``complexity equals anything" proposal can effectively probe such a complex internal structure, especially in scenarios with three or more horizons. 
It is a unique advantage brought by the flexibility of the ``complexity equals anything". 
In terms of the codimension-one gravitational observables, we adopt the suggestion given in Ref.~\cite{Jorstad:2023kmq}, where the extremal surface is driven deeper into the black hole interior by introducing higher-curvature corrections. 
 {Through this construction, we demonstrates that the codimension-one extremal surfaces can be extended into the spacetime regions inside the third (and, in general, odd-numbered) horizon. 
This implies that the generalized volume complexity is allowed to probe all the spacetime regions where the blackening factor $f(r)<0$. 
Importantly, this circumvents the limitation of the CV conjecture, whose result, for black holes with more than one horizon, is confined to the region between the outermost two horizons. 
In terms of the codimension-zero gravitational observables, 
we show that the corresponding extremal region can also extend into the spacetime region inside the third horizon, 
provided the outer curvature of its upper boundary is sufficiently large. 
Moreover, by exploiting the distinct physical characteristics of different gravitational observables,
we are able to successfully distinguish the various black hole interiors. }
At last, we generalize our framework to accommodate black hole solutions with an arbitrary number of horizons. 
This extension highlights the robustness of the “complexity equals anything” proposal and its potential as a general diagnostic tool for probing the internal structure of black holes with complex horizon structure. 

The structure of this paper is as follows: In Sec.~\ref{sec2}, we briefly introduce the Bardeen-AdS class black hole. 
In Sec.~\ref{sec3}, we review the ``complexity equals anything" proposal. 
In Sec.~\ref{sec4}, we calculate the codimension-one gravitational observables for the Bardeen-AdS class black hole with one, two, and three horizons respectively. 
Especially, in the case of three horizons, the flexibility of this holographic complexity proposal allows us to probe a wider spacetime region. 
In Sec.~\ref{sec5}, we explicitly calculate the codimension-zero gravitational observables for constant mean curvature surfaces. 
In Sec.~\ref{sec6}, we extend the above discussion to the AdS black hole with multiple horizons. 
Finally, we present our conclusions and discussions in Sec.~\ref{sec7}.

\section{The Bardeen-AdS class black hole}\label{sec2}
\begin{figure}[htbp]
    \centering
    \includegraphics[scale=0.15]{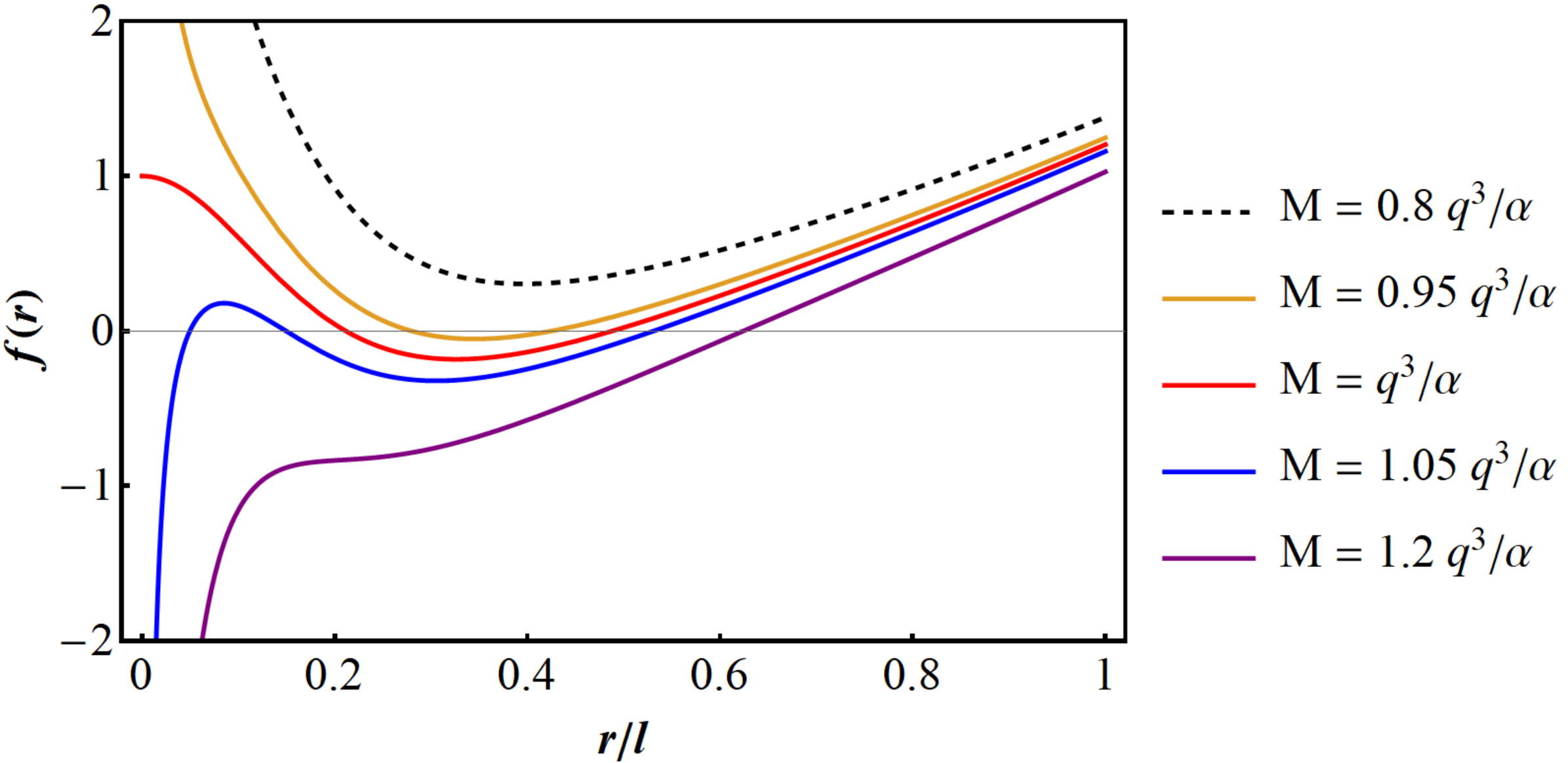}
    \caption[swx]{The blackening factor for the Bardeen-AdS class black hole with $l=5, \alpha=1, s=3, q=1.3$.
    The black dashed curve shows a naked singularity. 
    The brown curve represents that there is a black hole with two horizons and a timelike singularity. 
    The red curve shows that the black hole has still two horizons but the curvature singularity disappears. 
    The blue curve represents that there is a black hole with three horizons and a spacelike singularity. 
    The purple curve represents that there is a black hole with only one horizon and a spacelike singularity.
    }
   \label{fig1}
\end{figure}

In this section, we will introduce an example of multi-horizon black holes, which is the model we focus on. 
We consider the Einstein-AdS gravity coupled to nonlinear electrodynamics, whose action is given by \cite{Fan:2016hvf}
\begin{equation}
    \mathcal{I}=\frac{1}{16\pi G}\int d^{4}x \sqrt{-g} \left[R-\mathcal{L}(\mathcal{F},\alpha)+6l^{-2}\right],
\label{action}
\end{equation}
where $\mathcal{F}=F_{\mu\nu}F^{\mu\nu}$ is the square of the electromagnetic field tensor, 
and  {$\alpha > 0$ is the coupling constant with dimension of length squared of the nonlinear electromagnetic field, which measures the strength of electromagnetic field effects.
The Einstein and the nonlinear Maxwell equations are derived by varying Eq.~\eqref{action} with respect to $g_{\mu\nu}$ and $A_{\mu}$	respectively, i.e., 
\begin{align}
    R_{\mu\nu}-\frac{1}{2}g_{\mu\nu}R+\Lambda g_{\mu\nu}=&T_{\mu\nu},\\
    \nabla_{\mu}\left(\frac{\partial \mathcal{L}(\mathcal{F},\alpha)}{\partial \mathcal{F}} F^{\mu\nu}\right)=&0, 
\end{align} 
where 
\begin{equation}
   T_{\mu\nu}= -\frac{1}{2}\left(g_{\mu\nu}\mathcal{L}(\mathcal{F},\alpha)+4\frac{\partial \mathcal{L}(\mathcal{F},\alpha)}{\partial \mathcal{F}}\right). 
\end{equation} 
For the static spherically symmetric metric, we have
\begin{equation}
    \mathcal{L}(\mathcal{F},\alpha)=-2\left(\frac{f(r)'}{r}+\frac{f(r)-1}{r^{2}}+\Lambda\right).
\end{equation}
When considering the Bardeen-AdS class solution, its Lagrangian density $\mathcal{L}(\mathcal{F},\alpha)$ can be given by 
\begin{equation}
    \mathcal{L}=\frac{4s}{\alpha}\frac{(\alpha\mathcal{F})^{5/4}}{(1+\sqrt{\alpha\mathcal{F}})^{(1+s/2)}}, 
\end{equation}
where the square of the electromagnetic field tensor $\mathcal{F}$ is proportional to the square of the magnetic charge $Q_{\text{m}}$, that is  
\begin{equation}
    \mathcal{F}=\frac{2Q_{\text{m}}^{2}}{r^{4}}, 
\end{equation}
and $s>0$ is a dimensionless constant.}
Then, one can obtain the general two-parameter family black hole solution \cite{Fan:2016hvf,Wu:2024gqi}, i.e., 
\begin{align}
    ds^{2}&=-f(r)dt^{2}+\frac{dr^{2}}{f(r)}+r^{2}d\Omega^{2},\\
    A&=Q_{\text{m}}\cos{\theta}\,d\phi,\\
   f(r)&=1+\frac{r^{2}}{l^{2}}-\frac{2M}{r}+\frac{2q^{3}}{\alpha r}\left(1-\frac{r^{s}}{(r^{2}+q^{2})^{s/2}}\right), 
\end{align}
where $q$ is a free integration constant related to the magnetic charge, $Q_{\text{m}}=q^{2}/\sqrt{2\alpha}$. 
$M=M_{0}+M_{\text{em}}$ represents the ADM mass, which comprises two components: 
$M_{0}$, the Schwarzschild mass arising from the nonlinear self-interaction of the massless graviton, 
and $M_{\text{em}}=q^{3}/\alpha$, a charged term originating from the nonlinear interactions between the graviton and the photon. 
When $M_{0}=0$, i.e., $M=M_{\text{em}}=q^{3}/\alpha$, and $s = 3$, the above solution represents the Bardeen-AdS black hole, which has no curvature singularity at $r=0$~\cite{Ayon-Beato:2000mjt}. 
In this paper, we choose $s=3$ as an example, for which the blackening factor is shown in Fig.~\ref{fig1}. 
This model may have 0 to 3 horizons and 0 to 1 curvature singularity. We will only discuss the case where the event horizon exists.

\section{Complexity equals anything proposal}\label{sec3}
\begin{figure}[htbp]
    \centering
    \begin{subfigure}[b]{0.43\textwidth}
        \centering
		\includegraphics[scale=0.14]{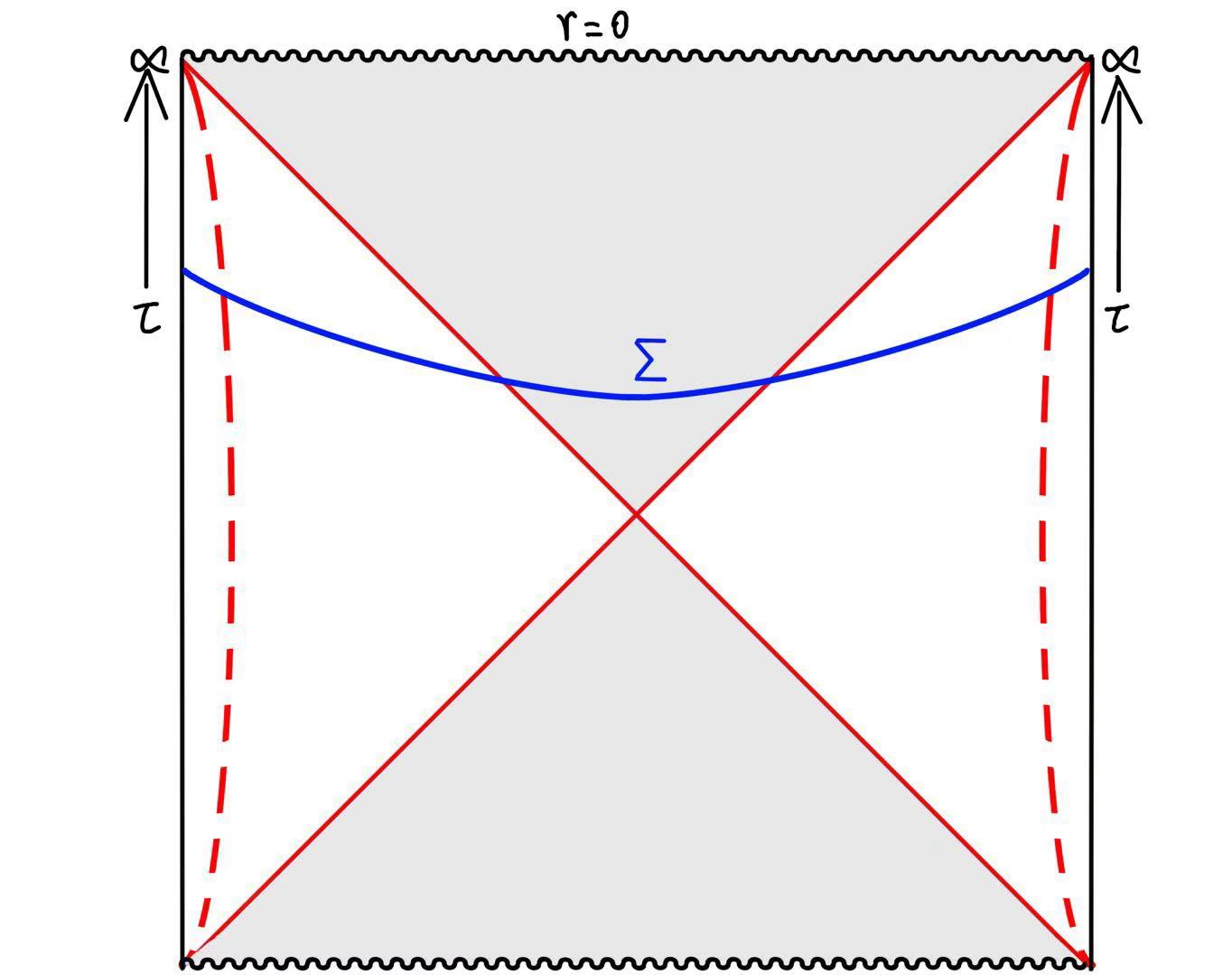}
        \caption{Codimension-one observable}
        \label{PDCo1}
    \end{subfigure}
    \hspace{0.04\textwidth}
    \begin{subfigure}[b]{0.43\textwidth}
        \centering
        \includegraphics[scale=0.14]{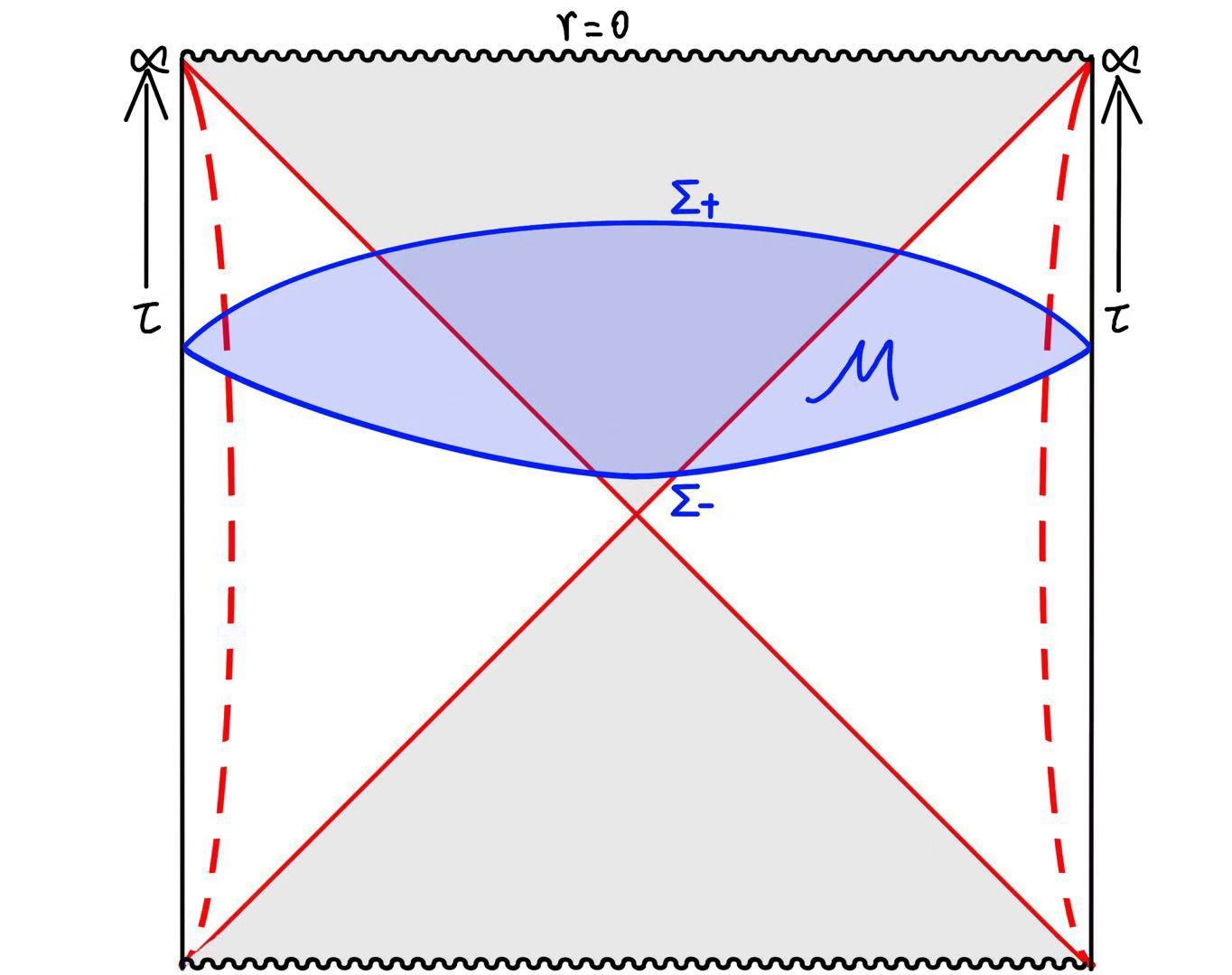}
        \caption{Codimension-zero observable}
        \label{PDCo0}
    \end{subfigure}
    \caption{(a). The codimension-one extremal surface $\Sigma$ anchored at the boundary time $\tau$. 
    (b). The codimension-zero extremal region $\Sigma_{+} \cup \mathcal{M} \cup \Sigma_{-}$ anchored at the boundary time $\tau$.}
   \label{PDCo1Co0}
\end{figure}

In this section, we will review the “complexity equals anything” proposal~\cite{Belin:2022xmt,Belin:2021bga} under the $(d+1)-$dimensional spherical symmetry metric as an example. 
The metric in Eddington-Finkelstein coordinate is given by
\begin{equation}
    ds^{2}=-f(r)dv^{2}+2dvdr+r^{2}d\Omega_{d-1}^{2},
\end{equation}
where $v=t+r_{*}(r)$ is the ingoing coordinate. 
Complexity measures the evolution of the TFD state entangling the CFTs on the left and right boundaries, i.e., 
 {
\begin{equation}
    |\text{TFD} \rangle =\frac{1}{\sqrt{Z(\beta,\mu)}}\sum_{n}e^{-\frac{\beta(E_{n}+\mu Q_{n})}{2}} |E_{n},Q_{n} \rangle_{L}\bigotimes |E_{n},Q_{n} \rangle_{R}. 
\end{equation}
where $Z(\beta, \mu)$ is the partition function of one copy of the CFT at temperature $\beta^{-1}$, 
$E_{n}$ are the energy eigenvalues of the single-sided Hamiltonian acting on one copy of the Hilbert space. 
$\mu$ is the chemical potential and $Q_{n}$ denotes the conserved charges. }
The ``complexity equals anything" proposal states that the complexity of the TFD state is dual to the gravitational observables on the extremal region in the bulk spacetime of an AdS two-sided black hole as shown in Fig.~\ref{PDCo1Co0}. 
In this paper, we will focus on the boundary time slices anchored at $t_{L}=t_{R}=\tau/2$ as usual. 

\subsection{Codimension-one observables}\label{sec3.1}
\begin{figure}[htbp]
    \centering
    \begin{subfigure}[b]{0.43\textwidth}
        \centering
		\includegraphics[scale=0.165]{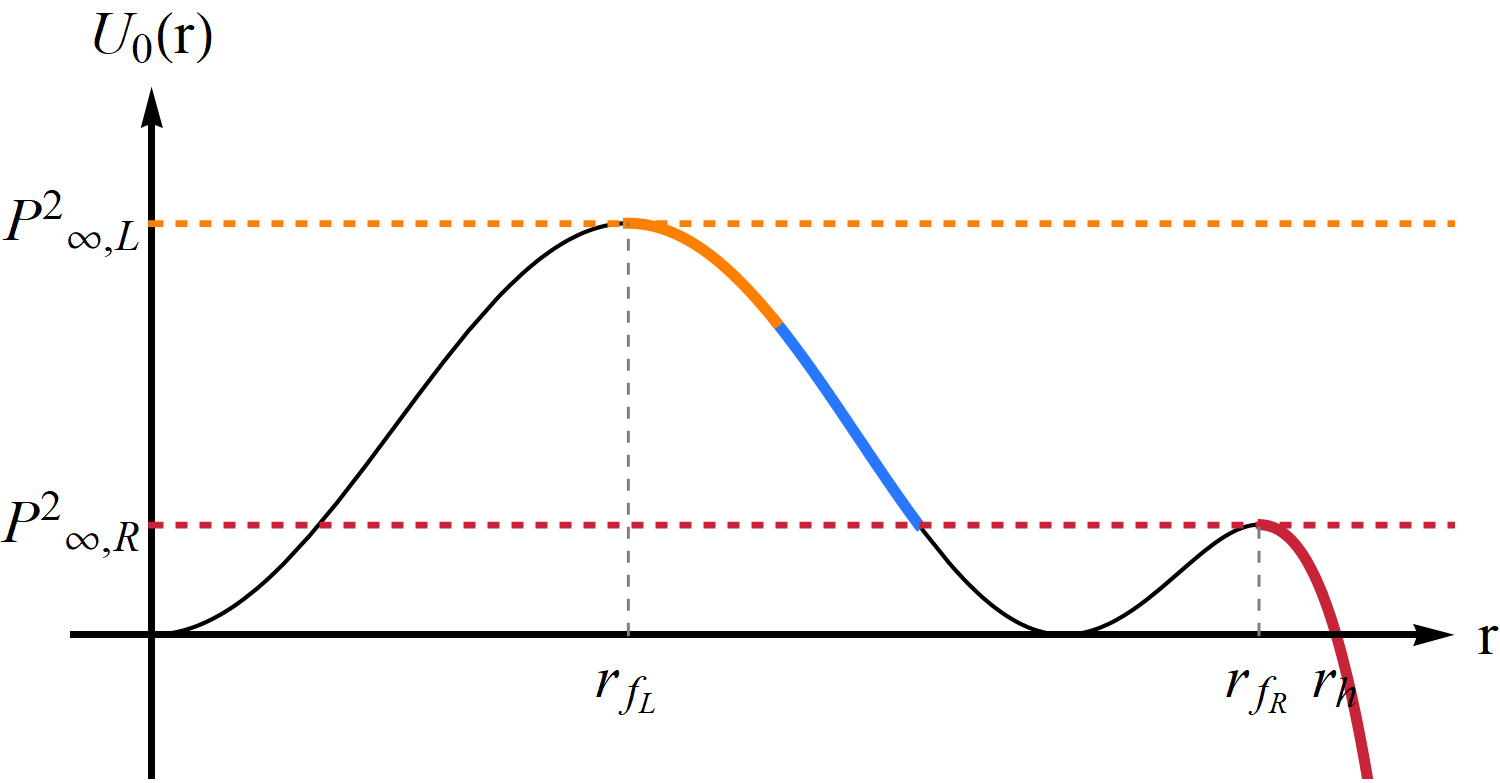}
        \caption{Effective potential}
        \label{UrEx}
    \end{subfigure}
    \hspace{0.08\textwidth}
    \begin{subfigure}[b]{0.43\textwidth}
        \centering
        \includegraphics[scale=0.12]{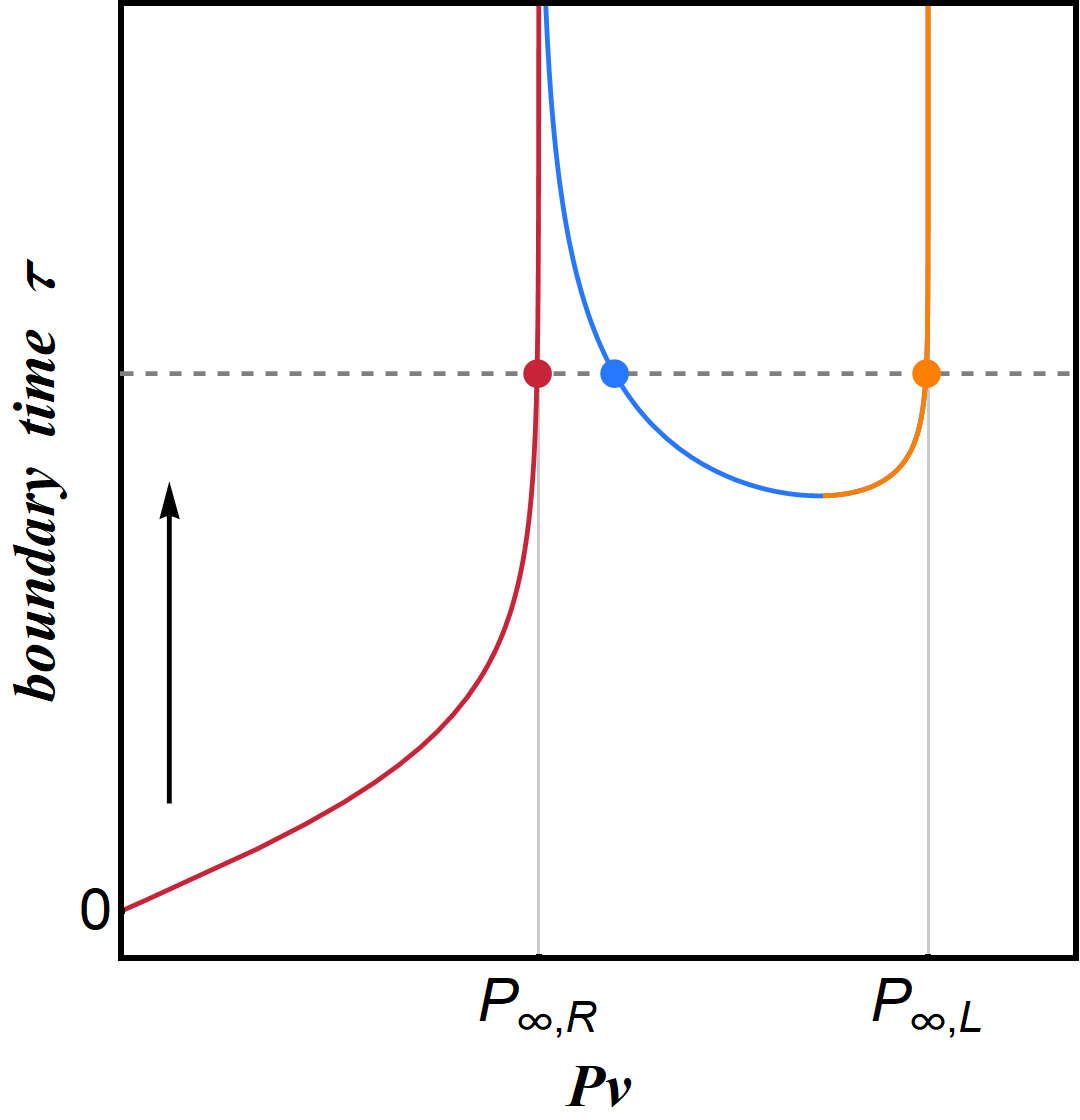}
        \caption{$P_{v}-\tau$}
        \label{PvtEx}
    \end{subfigure}
    \caption{(a) A common effective potential. Here we have two local maxima, one of them appears at $r=r_{f_{R}}$, the other one appears at $r=r_{f_{L}}$. The corresponding values of $P_{v}$ are $P_{\infty,R}$ and $P_{\infty,L}$, respectively. 
    (b) The relation between the conserved momentum $P_v$ and the boundary time $\tau$.}
   \label{ExEx}
\end{figure}

The expression of complexity with codimension-one observables is given by~\cite{Belin:2021bga}
\begin{equation}
    \mathcal{C}_{\text{gen}}[F_{1},\Sigma_{F_{2}}](\Sigma_{\text{CFT}})=\frac{1}{G_{\text{N}}l}\int_{\Sigma_{F_{2}}} \,d^{d} \sigma \,\sqrt{h}\, F_{1}(g_{\mu\nu};X^{\mu}),
\label{Cgen1}
\end{equation}
where $F_{1}$ and $F_{2}$ are two independent scalar functions, and $X^{\mu}$ is the embedding coordinate of the surface $\Sigma_{F_{2}}$. 
$\Sigma_{F_{2}}$ is determined by the extremization procedure, i.e.,
\begin{equation}
    \delta_{X}\left(\int_{\Sigma_{F_{2}}}d^{d}\sigma\sqrt{h}~F_{2}(g_{\mu\nu}; X^{\mu})\right)=0. 
\end{equation}
This proposal is commonly referred to as the ``generalized volume complexity''. 
~\\
~\\
\textbf{(a) The case of \boldmath $F_{1}=F_{2}$. }\\
When $F_{1}=F_{2}$, Eq.~\eqref{Cgen1} can be rewritten as
\begin{equation}
    \mathcal{C}_{\text{gen}}=\max _{\partial\Sigma=\Sigma_{\tau}}
    \left[\frac{1}{G_{\text{N}} l} \int_{\Sigma} d^{d} \sigma \sqrt{h}~a(r)\right], 
\label{CgenCo1}
\end{equation}
where $a(r)=F_{1}=F_{2}$ is a scalar function, and $\sigma$ denotes the coordinate on the surface $\Sigma$. 
Finding the extremal surface is analogous to solving a classical mechanics problem where the action and Lagrangian are identified as $S \sim \mathcal{C}_{\text{gen}} $ and $\mathcal{L} \sim \sqrt{h}\,a(r)$. 
Due to the symmetries of the spherical black hole, we can parametrize $\Sigma$ in terms of $(v(\sigma), r(\sigma), \Omega_{d-1})$, then, $\mathcal{C}_{\text{gen}}$ can be written as
\begin{equation}
    \mathcal{C}_{\text{gen}}=\frac{V_{d-1}}{G_{\text{N}}l}\int_{\Sigma}d\sigma r^{d-1}\sqrt{-f(r)\dot{v}^{2}+2\dot{v}\dot{r} }~a(r), 
\end{equation}
where $\dot{v}=dv/d\sigma$ and $\dot{r}=dr/d\sigma$. Since $\mathcal{C}_{\text{gen}}$ is invariant under the transformation $\sigma\to g(\sigma)$, we can choose a gauge condition as
\begin{equation}
    \sqrt{-f(r)\dot{v}^{2}+2\dot{v}\dot{r}}=a(r)r^{d-1}, 
\end{equation}
which allows us to derive the equation of motion, this problem can be interpreted as the motion of a classical non-relativistic particle in a potential, i.e., 
\begin{equation}
    \dot{r}^{2}+U_{0}(r)=P_{v}^{2}, 
\end{equation}
where the effect potential $U_{0}(r)$ is 
\begin{equation}
    U_{0}(r)=-f(r)a^{2}(r)r^{2(d-1)}. 
\end{equation}
For a given conserved momentum $P_{v}$, we can determine the minimum radius ($\dot{r}_{min}=0$) of the extremal surfaces and inversely solve for the boundary time at which it is anchored, i.e., 
\begin{equation}
    \tau=-2\int_{r_{\text{min}}}^{\infty}dr\frac{P_{v}}{f(r)\sqrt{P_{v}^{2}-U_{0}(r)}}.
    \label{tau}
\end{equation}
An important conclusion is that at late times, the extremal surfaces approach constant$-r$ surfaces, at which point the effective potential reaches a local maximum, i.e., 
\begin{equation}
    U_{0}(r_{f})= P_{\infty}^{2},\,\,\,\,U_{0}'(r_{f})=0,\,\,\,\,U_{0}''(r_{f})<0.
    \label{Co1Pinfty}
\end{equation}
We can give a typical example of the relation among the effective potential, the extremal surfaces and the boundary time, as shown in Fig,~\ref{ExEx}. 
For a given $\tau$, there are at most three different values of $P_{v}$. 
It means that there are at most three different extremal surfaces that can exist at the same boundary time. 
The red and yellow branches in Figs.~\ref{UrEx} and \ref{PvtEx} are the same as what we discussed in Eq.~\eqref{Co1Pinfty}. 
The blue branch is called the dipping branch, it was argued in Ref.~\cite{Belin:2022xmt} that it is not a good choice for discussing holographic complexity. 
After determining the extremal surfaces, Eq.~\eqref{CgenCo1} can be rewritten as
\begin{equation}
    \mathcal{C}_{\text{gen}}=-\frac{2V_{d-1}}{G_{\text{N}}l}\int_{r_{\text{min}}}^{r_{\infty}}\frac{U_{0}(r)}{f(r)\sqrt{P_{v}^{2}-U_{0}(r)}}dr.
\label{CgenF1eqF2}
\end{equation}
With Eqs.~\eqref{tau} and \eqref{CgenF1eqF2}, we can easily find that the evolution of complexity over time is proportional to $P_{v}$, i.e., 
\begin{equation}
    \frac{d\mathcal{C}_{\text{gen}}}{d\tau}=\frac{V_{d-1}}{G_{\text{N}}l}P_{v}.
\end{equation}
Obviously, when $\tau \rightarrow \infty$, we have
\begin{equation}
    \frac{d\mathcal{C}_{\text{gen}}}{d\tau}\bigg\vert _{\tau \rightarrow \infty}=\frac{V_{d-1}}{G_{\text{N}}l}P_{\infty }.
\end{equation}
~\\
\textbf{(b) The case of \boldmath $F_{1}\neq F_{2}$.}\\
The extremization process is analogous to the case where $F_{1}=F_{2}$ and the expression for complexity can be obtained directly, i.e.,
\begin{equation}
	\mathcal{C}_{F_{1},\Sigma_{F_{2}}}=-\frac{2V_{d-1}}{G_{\text{N}}l}\int_{r_{\text{min}}}^{\infty}\frac{\sqrt{U_{1}(r)U_{2}(r)}}{f(r)\sqrt{P_{v}^{2}-U_{2}(r)}}dr.
\end{equation}
This conclusion is consistent with what Ref.~\cite{Belin:2021bga} discussed. 
Its growth rate is only analytically solved at late times, i.e.,
\begin{equation}
	\lim_{\tau \to \infty} \frac{d\mathcal{C}_{F_{1},\Sigma_{F_{2}}}}{d \tau}=\frac{V_{d-1}}{G_{\text{N}}l}\sqrt{U_{1}(r_{f})}. 
\end{equation}
Regarding the difference between $F_{1}=F_{2}$ and $F_{1}\neq F_{2}$, Ref.~\cite{Jiang:2023jti} provided a detailed discussion.

\subsection{Codimension-zero observables}\label{sec3.2}
The expression of complexity with codimension-zero observables is given by~\cite{Belin:2022xmt}
\begin{align}
    \mathcal{C}_{\text{gen}}\left[G_{1},F_{1_{\pm}},\mathcal{M}_{G_{2},F_{2_{\pm}}}\right](\Sigma_{\text{CFT}})
    &=\frac{1}{G_{\text{N}}l}\int_{\Sigma_{+[G_{2},F_{2_{+}}]}} \,d^{d} \sigma \,\sqrt{h}\, F_{1_{+}}(g_{\mu\nu};X^{\mu}_{+})\notag\\
    &+\frac{1}{G_{\text{N}}l}\int_{\Sigma_{-[G_{2},F_{2_{-}}]}} \,d^{d} \sigma \,\sqrt{h}\, F_{1_{-}}(g_{\mu\nu};X^{\mu}_{-})\notag\\
    &+\frac{1}{G_{\text{N}}l^{2}}\int_{\mathcal{M}_{[G_{2},F_{2_{\pm}}]}} \,d^{d+1} x \,\sqrt{-g}\,G_{1}(g_{\mu\nu}). 
\end{align}
It is still need to obtain an extremal region, i.e.,
\begin{equation}
    \delta[W(\mathcal{M})]=0, 
    \label{extremize}
\end{equation}
where
\begin{align}
    W(\mathcal{M})=&\int_{\Sigma_+} \,d^{d} \sigma \,\sqrt{h}\, F_{2_{+}}(g_{\mu\nu};X^{\mu}_{+})
    +\int_{\Sigma_{-}} \,d^{d} \sigma \,\sqrt{h}\, F_{2_{-}}(g_{\mu\nu};X^{\mu}_{-})\notag\\
    &+\frac{1}{l}\int_\mathcal{M}\,d^{d+1} x \,\sqrt{-g}\,G_{2}(g_{\mu\nu}). 
\label{co0ES}
\end{align}
Thanks to the Stokes' theorem, 
Eq.~\eqref{extremize} can always be recast as two independent extremizations, i.e. 
\begin{equation}
   \delta_{\text{X}_{\pm}}\left[\int_{\Sigma_{\pm}}\,d^{d} \sigma \,\sqrt{h}\, F_{2_{\pm}}(g_{\mu\nu};X^{\mu}_{\pm}) \pm \widetilde{G}_{2}(g_{\mu\nu};X^{\mu}_{\pm})  \right]=0, 
\end{equation}
where $\widetilde{G}_{2}$ is obtained by Stokes' theorem, $viz$, 
\begin{align}
    \int_{\mathcal{M}} \, d^{d+1}x \, \sqrt{g} \, G_{2}(g_{\mu\nu}) =& l \int_{\partial{\mathcal{M}}=\Sigma_{+}\cup \Sigma_{-}} \, d^d \, \sigma \sqrt{h} \, \widetilde{G}_{2}(g_{\mu\nu};X^{\mu})\notag\\ 
=&l\left[\int_{\Sigma_{+}}\, d^d \, \sigma \sqrt{h} \, \widetilde{G}_{2}(g_{\mu\nu};X^{\mu}_{+})+\int_{\Sigma_{-}}\, d^d \, \sigma \sqrt{h} \, \widetilde{G}_{2}(g_{\mu\nu};X^{\mu}_{-})\right]. 
\end{align}
The process of extremization is similar to the codimension-one process, and we will again use the asymptotic AdS for spherically symmetric black holes as an example, i.e.,
 \begin{equation}
     ds^{2}=-f(r)dv^{2}+2dvdr+r^2d\Omega^{2}.
 \end{equation}
~\\
\textbf{(a) The case of \boldmath $G_{1}=G_{2},\,F_{1}=F_{2}$. }\\
Just as in the codimension-one case, our discussions can still start with a simple situation.
When $G_{1}=G_{2}=G$ and $F_{1}=F_{2}=F$, the complexity can be expressed as
\begin{align}
    \mathcal{C}_{\text{gen}}=\max_{\partial\Sigma=\Sigma_{\tau}} \Bigg[&\frac{1}{G_{\text{N}}l}\int_{\Sigma_{+}} \,d^{d} \sigma \,\sqrt{h}\, F_{+}(g_{\mu\nu};X^{\mu}_{+})+\frac{1}{G_{\text{N}}l}\int_{\Sigma_{-}} \,d^{d} \sigma \,\sqrt{h}\, F_{-}(g_{\mu\nu};X^{\mu}_{-})\notag\\
    +&\frac{1}{G_{\text{N}}l^{2}}\int_{\mathcal{M}} \,d^{d+1} x \,\sqrt{-g}\,G(g_{\mu\nu})\Bigg]. 
\end{align}
Anchoring at the boundary times and parametrize the surfaces, 
Stokes' theorem can still treat the codimension-zero observables in $\mathcal{M}$ as two independent contributions on the two boundaries $\Sigma_{\pm}$, that are 
\begin{equation}
    \mathcal{C}_{\text{gen}}^{\pm}(\tau)=\frac{V_{x}}{G_{\text{N}}l}\int_{\Sigma_{\pm}} d\sigma\left[r\sqrt{-f(r)\dot{v}^{2}+2\dot{v}\dot{r}}\,a_{\pm}(r)\mp \dot{v}b(r)\right],  
    \label{Cgen1=2}
\end{equation}
where $b(r)$ is obtained by 
\begin{equation}
    \sqrt{-g}G_{1}(g_{\mu\nu})=G_{1}(r)r^{d-1}\equiv l\frac{\partial b(r)}{\partial r}.
    \label{b(r)}
\end{equation}
This problem can still be viewed as two independent classical mechanics problems with the following two Lagrangians: 
\begin{equation}
    \mathcal{L}_{\pm}:=r^{d-1}\sqrt{-f(r)\dot{v}^{2}+2\dot{v}\dot{r}}\,a_{\pm}(r)\mp\dot{v}b(r).
\label{co0L}
\end{equation}
Choosing the gauge condition as
\begin{equation}
    r^{d-1}\,a_{\pm}(r)=\sqrt{-f(r)\dot{v}^{2}+2\dot{v}\dot{r}},
\end{equation}
the conserved generalized momentum and the extremality conditions can be given by 
\begin{align}
    P_{v}^{\pm}&=\dot{r}-f(r)\dot{v}\mp\dot{b}(r),\\
    \dot{r}&=\sqrt{(P_{v}^{\pm}\pm b(r))^{2}+f(r)a_{\pm}^{2}(r)r^{2(d-1)}},\\
    \dot{v}&=\frac{\dot{r}-P_{v}^{\pm}\mp b(r)}{f(r)}.
\end{align}
Then the equation of motion can be defined as
\begin{align}
    \dot{r}+\mathcal{U}_{\pm}(P_{v}^{\pm},r)=0,
\end{align}
where
\begin{align}
    \mathcal{U}_{\pm}(P_{v}^{\pm},r)=&U_{0}(r)-(P_{v}^{\pm}\pm b(r))^{2}, \\
    U_{0}(r)=&-f(r)a_{\pm}^{2}(r)r^{2(d-1)}.
    \label{Upmrf}
\end{align}
When $r=r_{\text{min}}$, $\mathcal{U}_{\pm}(P_{v}^{\pm},r_{\text{min}})=0$. 
For a given $P_{v}^{\pm}$, the minimal radius can be determined by
\begin{equation}
    P_{v}^{\pm} = \zeta \sqrt{U_{0}(r_{\text{min}})} \mp b(r_{\text{min}}), 
    \label{Pvrmin}
\end{equation}
where $\zeta=\pm$. 
Alternatively, for a given $r_{\text{min}}$, the conserved momentum can be calculated. 
By inversely solving the boundary time, the time evolution of $\mathcal{C}_{\text{gen}}$ can be easily obtained, $viz$, 
\begin{align}
    \tau&=-2\int_{r_{\text{min}}}^{\infty}\,dr\,\frac{P_{v}^{\pm}\pm b(r)}{f(r)\sqrt{(P_{v}^{\pm}\pm b(r))^{2}-U_{0}(r)}}, \\
    \frac{d}{d\tau}&\mathcal{C}_{\text{gen}}(\tau)=\frac{V_{d-1}}{G_{\text{N}}l}\left(P_{v}^{+}(\tau)+P_{v}^{-}(\tau)\right),
\end{align}
where $\mathcal{C}_{\text{gen}}=\mathcal{C}_{\text{gen}}^{+}(\tau)+\mathcal{C}_{\text{gen}}^{-}(\tau)$. 
$P_{v}^{\pm}(\tau)$ will approach constant values at $\tau \to \infty$, where the extremal surfaces tends to constant$-r$ surfaces, i.e. $r=r_{f}$. 
The $r_{f}$ must correspond to a local maximum of the effective potential, that is
\begin{equation}
    \mathcal{U}_{\pm}(P_{\infty}^{\pm},r_{f})=0,\,\,\,\partial_{r}\mathcal{U}_{\pm}(P_{\infty}^{\pm},r_{f})=0,\,\,\,\partial_{r}^{2}\mathcal{U}_{\pm}(P_{\infty}^{\pm},r_{f})< 0. 
\label{tauinfty}
\end{equation}
where $P_{\infty}^{\pm}=\lim\limits_{\tau \to \zeta \infty} P_{v}^{\pm}(\tau)$. The full derivation details were given in Ref.~\cite{Belin:2022xmt}. \\
~\\
\textbf{(b) The case of \boldmath $G_{1}\neq G_{2},\,F_{1}\neq F_{2}$. }\\
The whole extremization for the case of $G_{1}\neq G_{2},\,F_{1}\neq F_{2}$ is similar to the case of $G_{1}=G_{2},\,F_{1}=F_{2}$, and the expression of complexity can be obtained directly: 
\begin{equation}
    \mathcal{C}_{\text{gen}}^{\pm}(\tau)=-\frac{2V_{d-1}}{G_{\text{N}}l}
    \int_{r_{\text{min}}}^{\infty} dr \left[\frac{\sqrt{U_{1}(r)U_{2}(r)}-b_{1}(r)(b_{2}(r)\pm P_{v}^{\pm})}{f(r)\,\sqrt{(P_{v}^{\pm}\pm b_{2}(r))^{2}-U_{2}(r)}}\right], 
\end{equation}
where $U_{1,2}(r)=-f(r)a^{2}_{1,2}(r)r^{2(d-1)},\,P_{v}^{\pm}=\zeta \sqrt{U_{2}(r_{\text{min}})}\mp b_{2}(r_{\text{min}})$. 
The late time growth can be obtained by 
\begin{equation}
    \lim_{\tau \to \zeta \infty}  \frac{d}{d\tau}\mathcal{C}_{\text{gen}}(\tau)=\frac{V_{d-1}}{G_{\text{N}}l}(\mathcal{P}_{\infty}^{+}(\tau)+\mathcal{P}_{\infty}^{-}(\tau)).
\end{equation}
where $\mathcal{P}_{\infty}^{\pm}\equiv\zeta \sqrt{U_{1}(r_{f})} \mp b_{1}(r_{f})$. 
~\\

When $G_{2}$ and $F_{2}$ are constants, Eq.~\eqref{co0ES} can be rewritten as 
\begin{equation}
    W(\mathcal{M})=\frac{1}{G_{\text{N}}l}\left[\alpha_{+}\int_{\Sigma_{+}}\,d^{d}\sigma\,\sqrt{h}+\alpha_{-}\int_{\Sigma_{-}}\,d^{d}\sigma\,\sqrt{h}+\frac{\alpha_{B}}{l}\int_{\mathcal{M}}\,d^{d+1}x\,\sqrt{-g}\right].
\end{equation}
In this case, the two boundaries of the extremal region $\mathcal{M}$ becomes two constant mean curvature (CMC) slices~\cite{CMC1,CMC2}: 
\begin{equation}
    K_{\Sigma_{+}}=-\frac{\alpha_{B}}{\alpha_{+}l},~~~~~K_{\Sigma_{-}}=\frac{\alpha_{B}}{\alpha_{-}l}, 
\end{equation} 
where $K_{\Sigma_{\pm}}$ are the traces of the extrinsic curvature tensors $K_{\mu\nu}^{\pm}\equiv h^{\alpha}_{\mu}\bigtriangledown_{\alpha} n_{\nu}^{\pm}$, 
and the timelike normal vectors $n^{\mu}_{\pm}$ is chosen to be future-directed for the spacelike surfaces $\Sigma_{\pm}$.
We can verify this conclusion at late times $(r=r_{f})$:
\begin{align}
    n^{\mu}_{\pm}\Big|_{r=r_{f}}=(0,\frac{1}{\sqrt{-f(r_{f})}},0,0),\\
    K_{\Sigma_\pm}\Big|_{r=r_{f}}=\frac{4f(r_{f})+r_{f}f'(r_{f})}{2r_{f}\sqrt{-f(r_{f})}}, \label{ExCur}
\end{align}
where $r_{f}$ can be obtained by Eq.~\eqref{tauinfty}.
Returning to the extremizations, the bulk term can still be equivalent to contributions at the boundaries, i.e., 
\begin{align}
    W(\mathcal{M})_{\text{CMC}}=&\int_{\Sigma_{+}}\,d\sigma\,\left[r^{d-1}\sqrt{-f(r)\dot{v}^{2}+2\dot{v}\dot{r}}\, - \frac{\alpha_{B}}{d \alpha_{+} l}\dot{v}r^{d} \right]\notag\\
    +&\int_{\Sigma_{-}}\,d\sigma\,\left[r^{d-1}\sqrt{-f(r)\dot{v}^{2}+2\dot{v}\dot{r}}\, + \frac{\alpha_{B}}{d \alpha_{-} l}\dot{v}r^{d}\right], 
    \label{WCMC}
\end{align}
where 
\begin{equation}
    a_{2}(r)=1,\,\,\,\,b_{2}(r)=\frac{\alpha_{B}}{d\alpha_{\pm}l}r^{d}.
    \label{br}
\end{equation}
Focusing on one of the extremal surfaces, that is, setting $G_{1}$ and either $F_{1_{+}}$ or $F_{1_{-}}$ to zero, 
the complexity and its late time growth can be obtained as 
\begin{align}
    \mathcal{C}_{\text{gen}}^{\pm}&=\frac{1}{G_{\text{N}}l}\int_{\Sigma_{\pm}(\alpha_{\pm},\alpha_{B})}\,d^{d}\sigma\,\sqrt{h}\,a(r),
    \label{CCMCpm}\\
    \lim_{\tau\to\infty}\frac{d}{d\tau} \mathcal{C}_{\text{gen}}^{\pm}&=
    \frac{V_{d-1}}{G_{\text{N}}l}\,r_{f}^{d-1}\,a(r_{f})\,\sqrt{-f(r_{f})},
\label{dCCMClate}
\end{align}
where $a(r)$ is a dimensionless scalar function. This is often referred to as the ``generalized complexity".

\section{Generalized volume complexity for Bardeen-AdS class black hole}\label{sec4}
As discussed in Sec.\ref{sec2}, the Bardeen-AdS class black hole can have up to three horizons. 
In this section, we will briefly compute the generalized volume complexity for this model with one or two horizons, 
and discuss the three-horizon case in detail.

\subsection{The one-horizon case}
\begin{figure}
    \centering
    \includegraphics[scale=0.16]{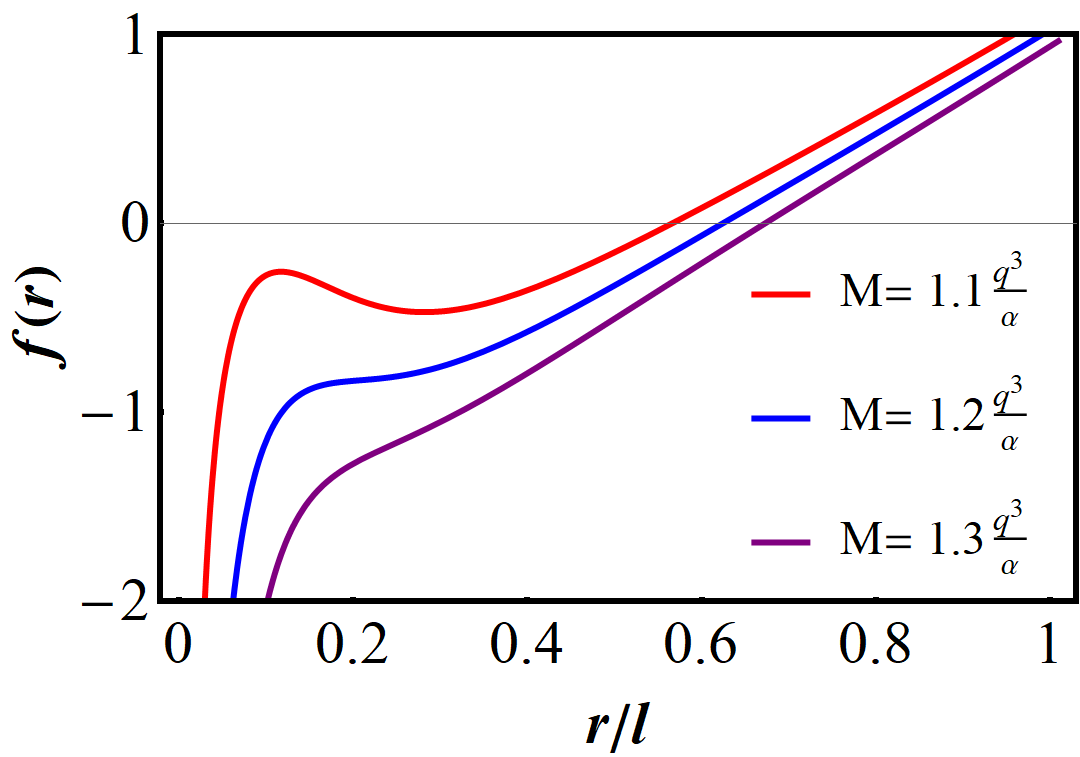}
    \caption{The blackening factor $f(r)$ for the Bardeen-AdS class black hole with $ l = 5,\,\alpha = 1,\,s = 3,\, q = 1.3 $, for which there is only one zero point. }
    \label{fr1h}
\end{figure}

\begin{figure}[htbp]
    \centering
    \begin{subfigure}[htbp]{0.45\textwidth}
        \centering
		\includegraphics[scale=0.16]{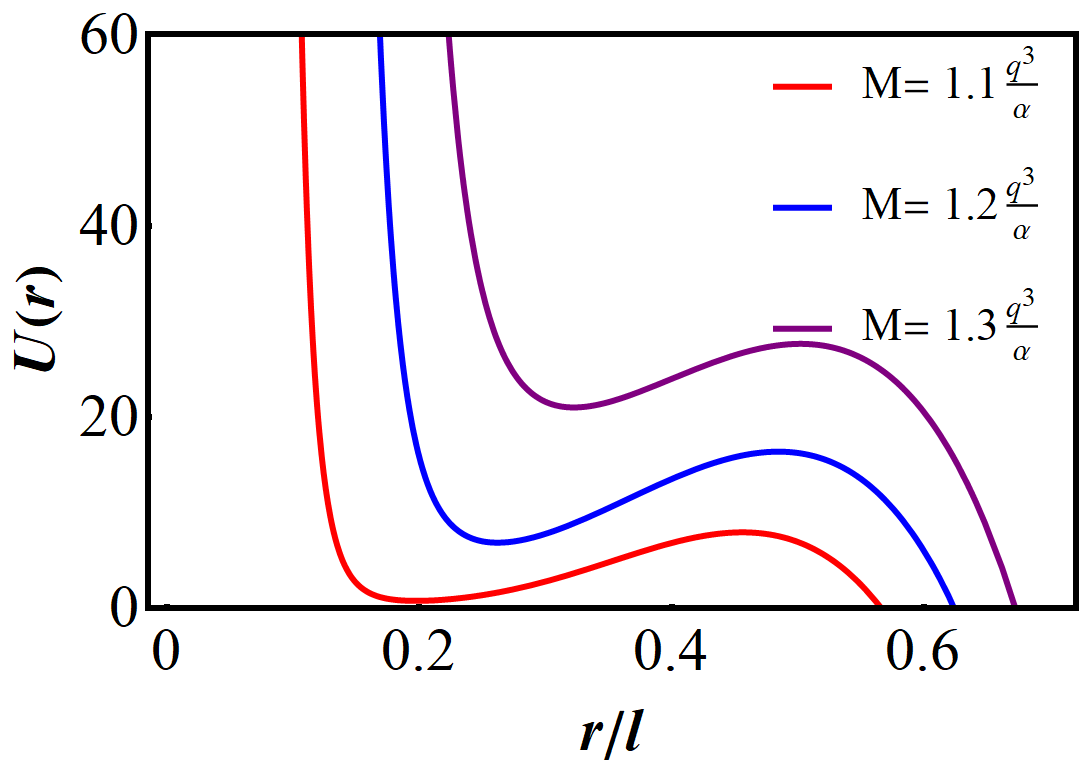}
        \caption{ $a(r)=1+\lambda\,l^{4} C^{2}$}
        \label{Ur11}
    \end{subfigure}
    \hspace{0.05\textwidth}
    \begin{subfigure}[htbp]{0.45\textwidth}
        \centering
        \includegraphics[scale=0.16]{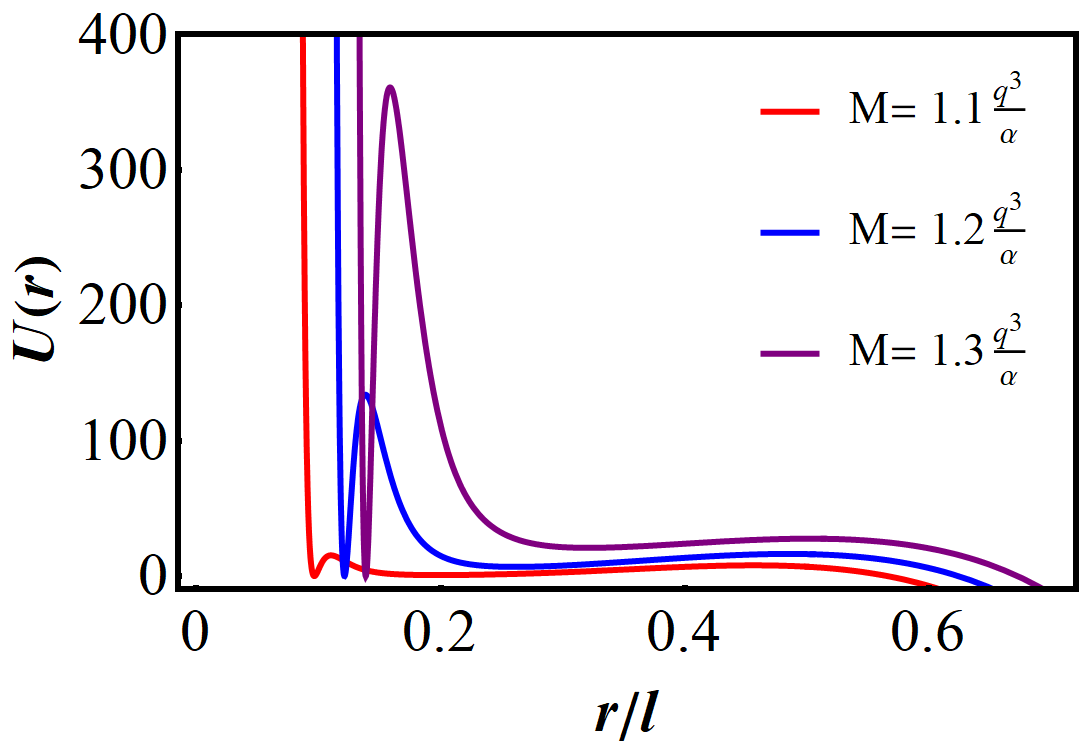}
        \caption{$a(r)=1+\lambda_{1} l^{4} C^{2}-\lambda_{2} l^{8} C^{4}$}
        \label{Ur12}
    \end{subfigure}
    \caption{The effective potentials with $l = 5, \alpha = 1, s = 3, q = 1.3, \lambda = \lambda_{1} = 10^{-3}, \lambda_{2} = 10^{-8} $.
    (a) The effective potentials with only one local maximum. 
    (b) The effective potentials with two local maxima. 
    }
   \label{Ur1H}
\end{figure}

\begin{figure}[htbp]
    \centering
    \begin{subfigure}[htbp]{0.45\textwidth}
        \centering
		\includegraphics[scale=0.16]{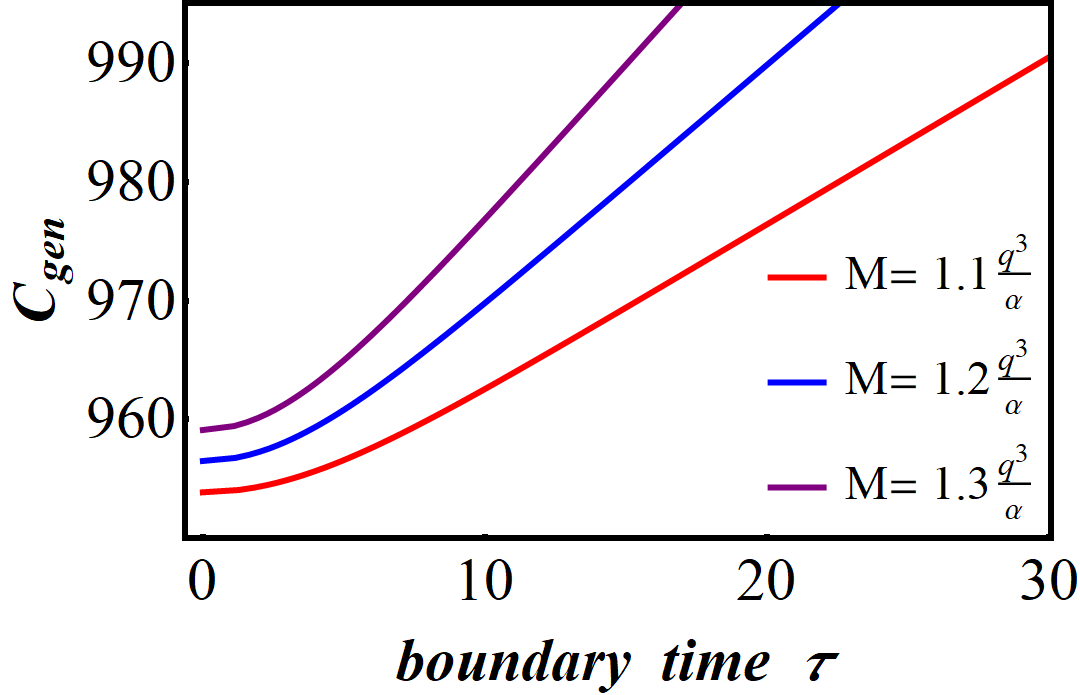}
        \caption{$a(r)=1+\lambda\,l^{4} C^{2}$}
        \label{Cgen1h1f}
    \end{subfigure}
    \hspace{0.03\textwidth}
    \begin{subfigure}[htbp]{0.45\textwidth}
        \centering
        \includegraphics[scale=0.16]{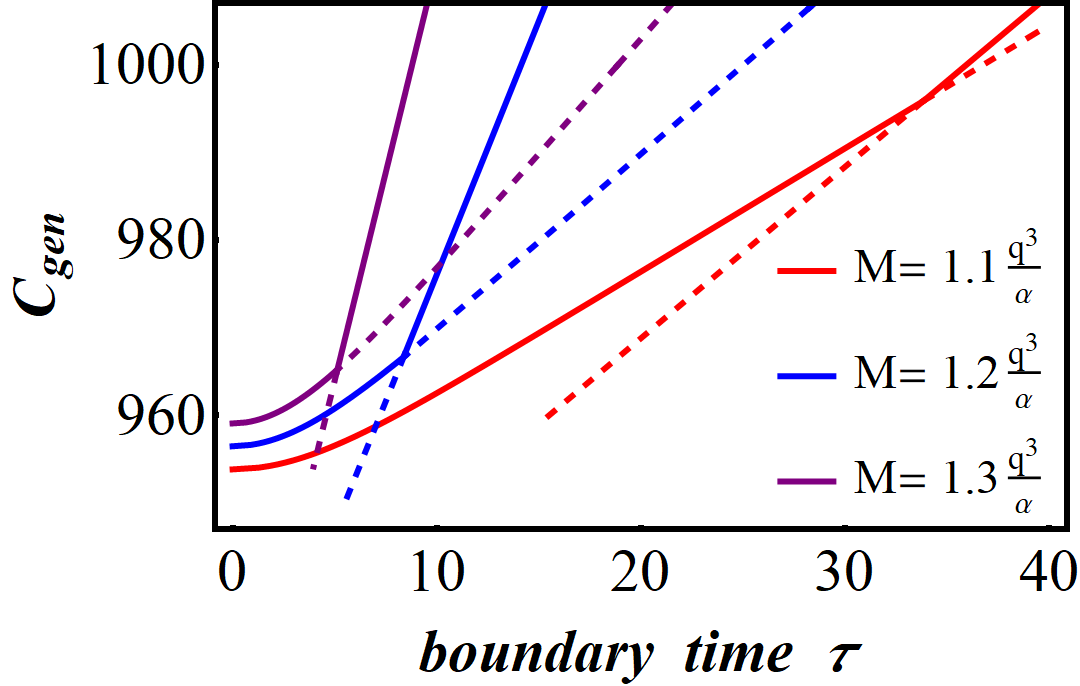}
        \caption{$a(r)=1+\lambda_{1} l^{4} C^{2}-\lambda_{2} l^{8} C^{4}$}
        \label{Cgen1h2f}
    \end{subfigure}
    \caption{Time evolution of the generalized volume complexity with $l = 5, \alpha = 1, s = 3, q = 1.3, \lambda=\lambda_{1} = 10^{-3}, \lambda_{2} = 10^{-8}$.
    (a) $a(r)=1+\lambda\,l^{4} C^{2}$. The complexity grows linearly at late times, and the growth rate increases with the increase of $M$.
    (b) $a(r)=1+\lambda_{1} l^{4} C^{2}-\lambda_{2} l^{8} C^{4}$. 
    The dashed line shows the non-maximum generalized volume of the extremal surfaces, which are not considered suitable as holographic duals of the complexity. 
    The growth rate of the complexity is abrupt while the complexity itself is still continuous.}
   \label{Cgen1h}
\end{figure}

We begin our discussion with the one-horizon case, where the black hole has a spacelike singularity. The corresponding blackening factor $f(r)$ for this case is illustrated in Fig.~\ref{fr1h}.
Considering the generalized volume complexitiy, we can construct the scalar function $a(r)$ using the square of the Weyl tensor $C^{2}\equiv C_{\mu\nu\sigma\rho}C^{\mu\nu\sigma\rho}$ and a subleading term $C^{4}\equiv (C_{\mu\nu\sigma\rho}C^{\mu\nu\sigma\rho})^2$, i.e., 
\begin{align}
    a(r)=&1+\lambda l^{4} C^{2}, \label{ar1}\\
    \text{or}\,\,\,\,\,\,\,\,\,\,a(r)=&1+\lambda_{1} l^{4} C^{2}-\lambda_{2} l^{8} C^{4}, \label{ar2}
\end{align}
where $\lambda,\,\lambda_{1},\,\lambda_{2}$ are dimensionless parameters, and $l$ is the AdS radius. 
The square of the Weyl tensor $C^{2}$ can be given by
\begin{equation}
    C^{2}=\frac{(2-2f(r)+2rf'(r)-r^{2}f''(r))^{2}}{3 r^{4}}. 
    \label{WeylTensor2}
\end{equation}
Selecting the parameters given in Fig.~\ref{fr1h}, we compute the effective potentials from Eqs.~\eqref{ar1} and \eqref{ar2}, respectively. 
By adjusting the value of the dimensionless parameters to ensure that the effective potentials exhibit at least one positive local maximum, 
we can obtain the images of such potential functions as shown in Fig.~\ref{Ur1H}. 
The higher curvature term allows the effective potential to have more than one local maximum, enabling us to obtain more extremal surfaces. 
Further, the generalized volume complexity~\eqref{CgenF1eqF2} can be computed by these two constructions, respectively, as shown in Fig.~\ref{Cgen1h}.
It is important to note that the value of $\mathcal{C}_{\text{gen}}$ itself lacks physical significance, as it is dependent on the position of the cutoff. 
The part we are interested in is its growth rate, that is, $d\mathcal{C}_{\text{gen}}/d\tau$, and the relative value of the generalized volumes for different extremal surfaces under the same cutoff. 
When the effective potential has multiple local maxima, the complexity may grow in a non-smooth way as shown in Fig.~\ref{Cgen1h2f}. 
Such behavior is widespread, and the moment when the phase transition occurs is called the turning time. 
We have discussed this phenomenon in detail in Ref.~\cite{Jiang:2023jti}. 
A clear physical explanation for the phenomenon of multiple local maxima in the effective potential was recently provided by Caceres et al.~\cite{Caceres:2025myu}. 
In particular, the circuit complexity of multiple local optima using a finite qubit system was simulated, offering insights into how this behavior can be understood from a boundary perspective. 

\subsection{The two-horizon case}
\begin{figure}
    \centering
    \includegraphics[scale=0.15]{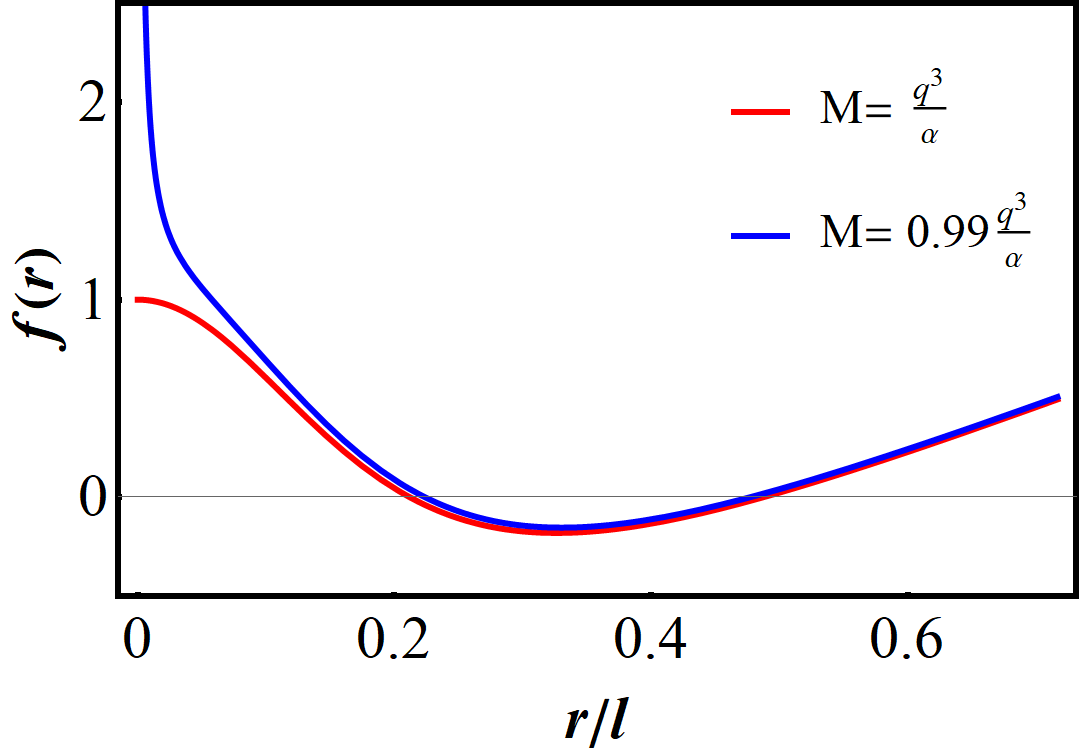}
    \caption{The blackening factor $f(r)$ for the Bardeen-AdS class black hole with $ l = 5,\,\alpha = 1,\,s = 3,\, q = 1.3 $, for which there are two zero points.}
    \label{fr2h}
\end{figure}

\begin{figure}[htbp]
    \centering
    \begin{subfigure}[b]{0.45\textwidth}
        \centering
		\includegraphics[scale=0.15]{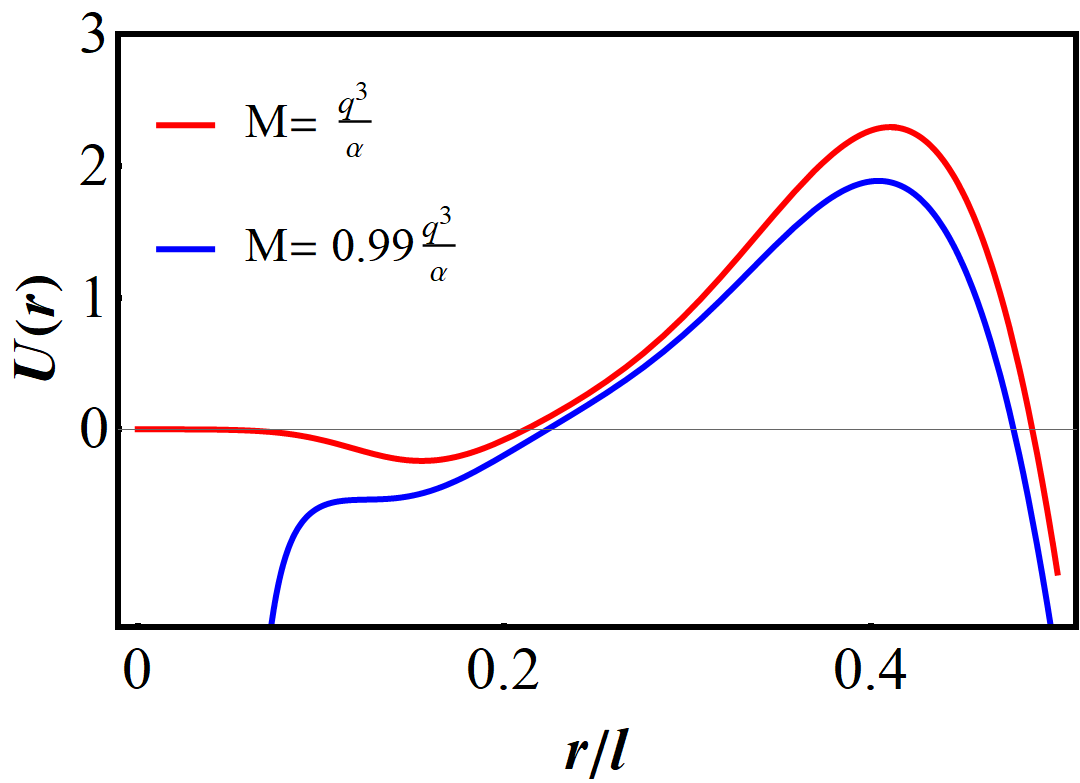}
        \caption{$a(r)=1+\lambda\,l^{4} C^{2}$}
        \label{Ur21}
    \end{subfigure}
    \hspace{0.03\textwidth}
    \begin{subfigure}[b]{0.45\textwidth}
        \centering
        \includegraphics[scale=0.15]{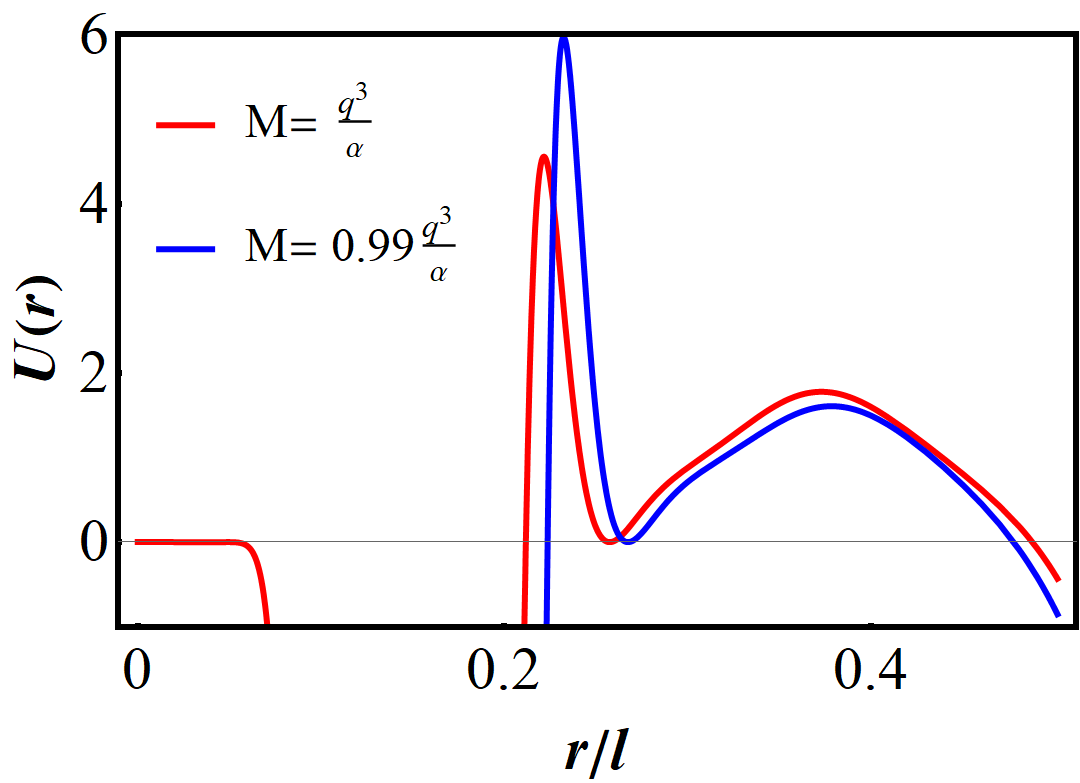}
        \caption{$a(r)=1+\lambda_{1} l^{4} C^{2}-\lambda_{2} l^{8} C^{4}$}
        \label{Ur22}
    \end{subfigure}
    \caption{The effective potentials with $l = 5, \alpha = 1, s = 3, q = 1.3, \lambda = \lambda_{1} = 10^{-3}, \lambda_{2} = 2.5*10^{-4} $. 
    At $r\to0$, the behavior of $U(r)$ (divergent or finite) depends on the existence of the curvature singularity. 
    (a) The effective potentials have only one local maximum.
    (b) The effective potentials have two local maxima.
    }
   \label{Ur2H}
\end{figure}

\begin{figure}[htbp]
    \centering
    \begin{subfigure}[b]{0.45\textwidth}
        \centering
		\includegraphics[scale=0.15]{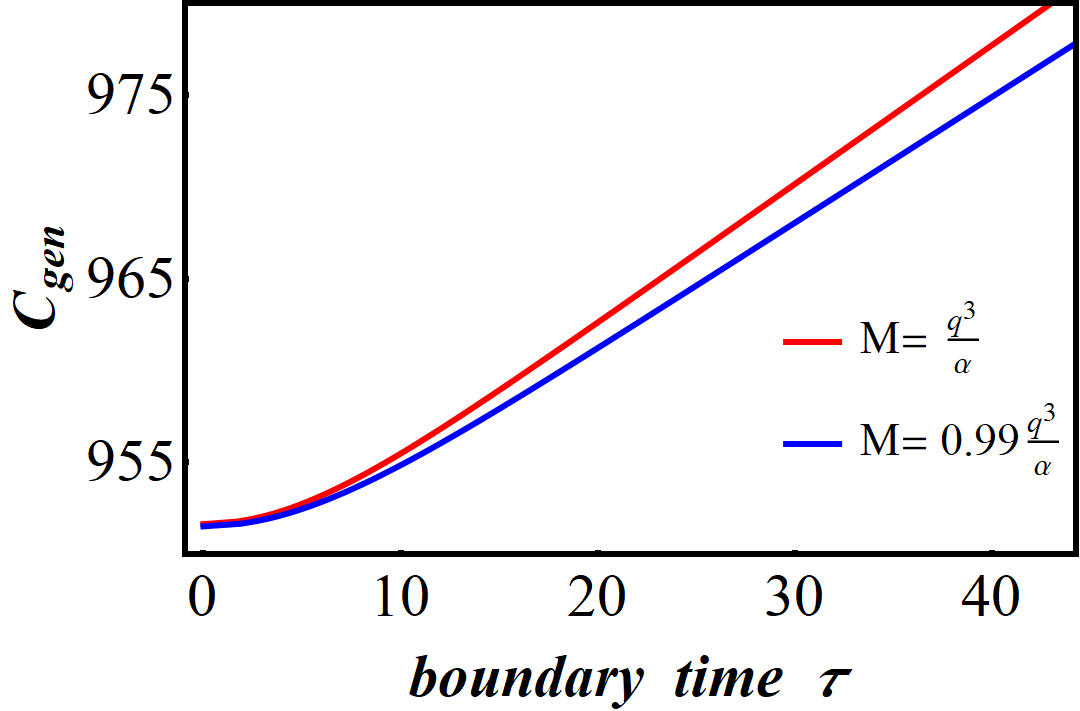}
        \caption{$a(r)=1+\lambda\,l^{4} C^{2}$}
        \label{Cgen2h1f}
    \end{subfigure}
    \hspace{0.03\textwidth}
    \begin{subfigure}[b]{0.45\textwidth}
        \centering
        \includegraphics[scale=0.15]{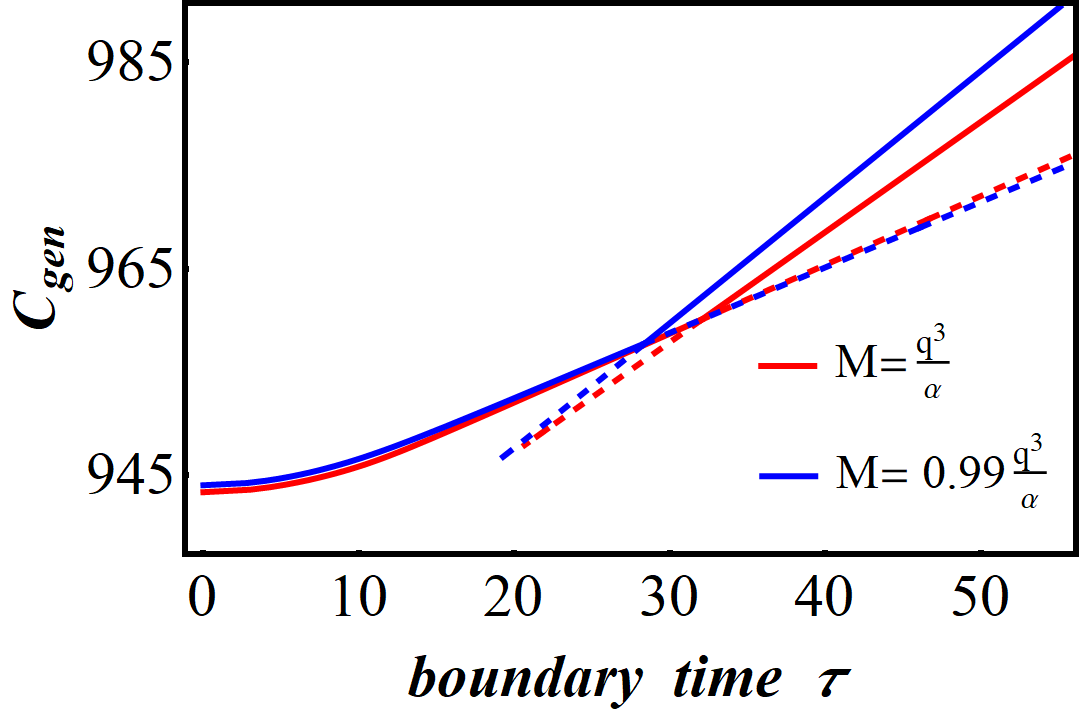}
        \caption{$a(r)=1+\lambda_{1} l^{4} C^{2}-\lambda_{2} l^{8} C^{4}$}
        \label{Cgen2h2f}
    \end{subfigure}
    \caption{Time evolution of the generalized volume complexity with $l = 5, \alpha = 1, s = 3, q = 1.3, \lambda = \lambda_{1} = 10^{-3}, \lambda_{2} = 2.5 \times 10^{-4} $.
    (a) The complexity grows linearly at late times, and the growth rate increases with the increase of $M$.
    (b) The growth rate of the complexity is abrupt while the complexity itself is still continuous.
    }
   \label{Cgen2h}
\end{figure}

In this subsection, we consider the two-horizon case, with the corresponding blackening factor $f(r)$ shown in Fig.~\ref{fr2h}. When $M=q^{3}/\alpha$, the timelike singularity can be eliminated. 
We still use the scalar function constructions given in Eqs.~\eqref{ar1} and \eqref{ar2}. The resulting effective potentials are illustrated in Fig.~\ref{Ur2H}. 
We observe that even though the singularity causes the effective potential to diverge at $r=0$, 
the complexity remains insensitive even when the timelike singularity is removed.
This is because the effective potential remains non-positive in the spacetime regions where $f(r) > 0$. 
The time evolution for the generalized volume complexity is shown in Fig.~\ref{Cgen2h}. 
The results indicate that the complexity growth rate is related to the black hole parameters. 
However, it fails to probe the existence or absence of the timelike singularity. 

\subsection{The three-horizon case}\label{4.3}
\begin{figure}[htbp]
    \centering
    \begin{subfigure}[b]{0.45\textwidth}
        \centering
		\includegraphics[scale=0.2]{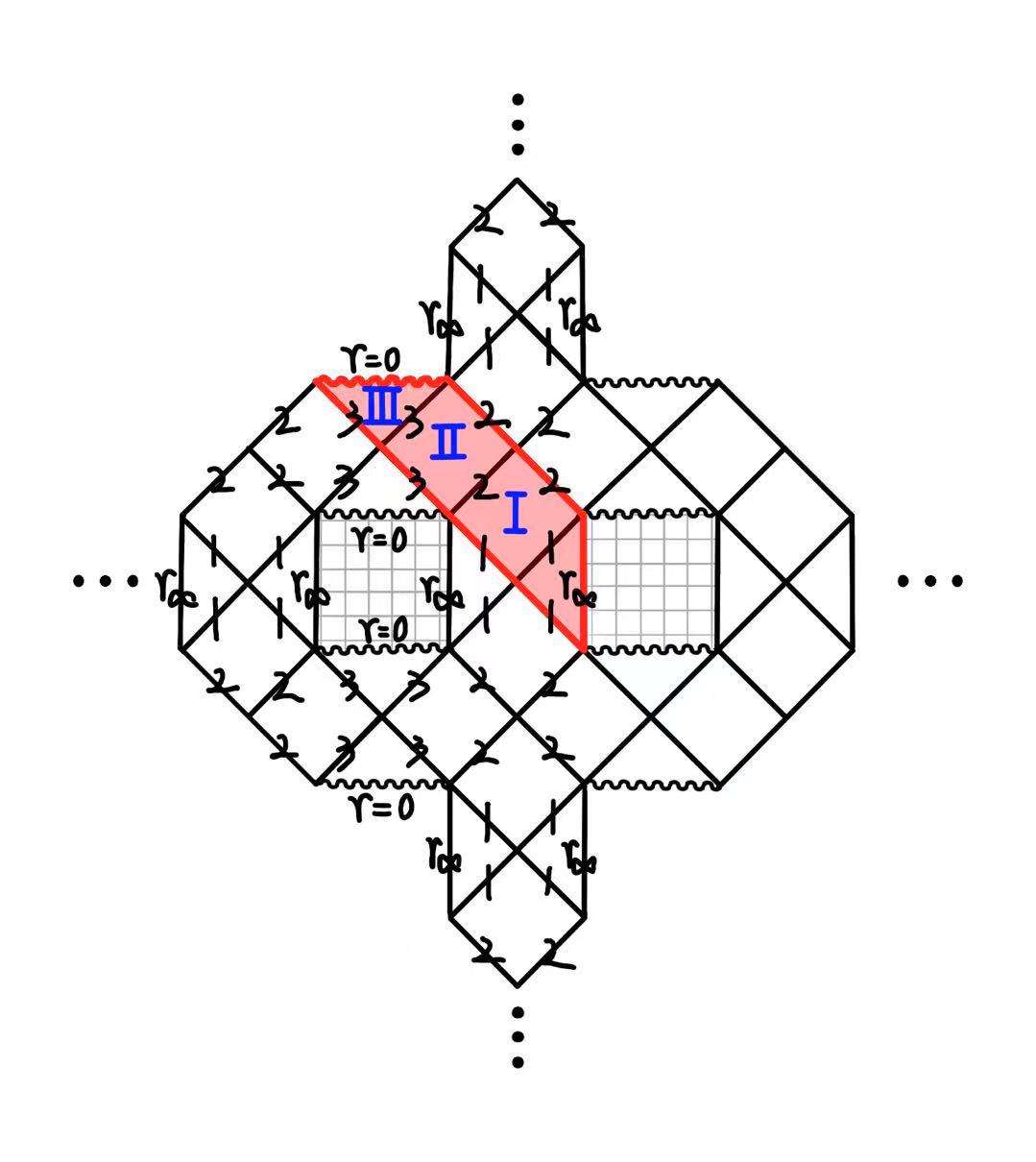}
        \caption{The Penrose diagram}
        \label{PenroseDiagram3}
    \end{subfigure}
    \hspace{0.03\textwidth}
    \begin{subfigure}[b]{0.45\textwidth}
        \centering
        \includegraphics[scale=0.16]{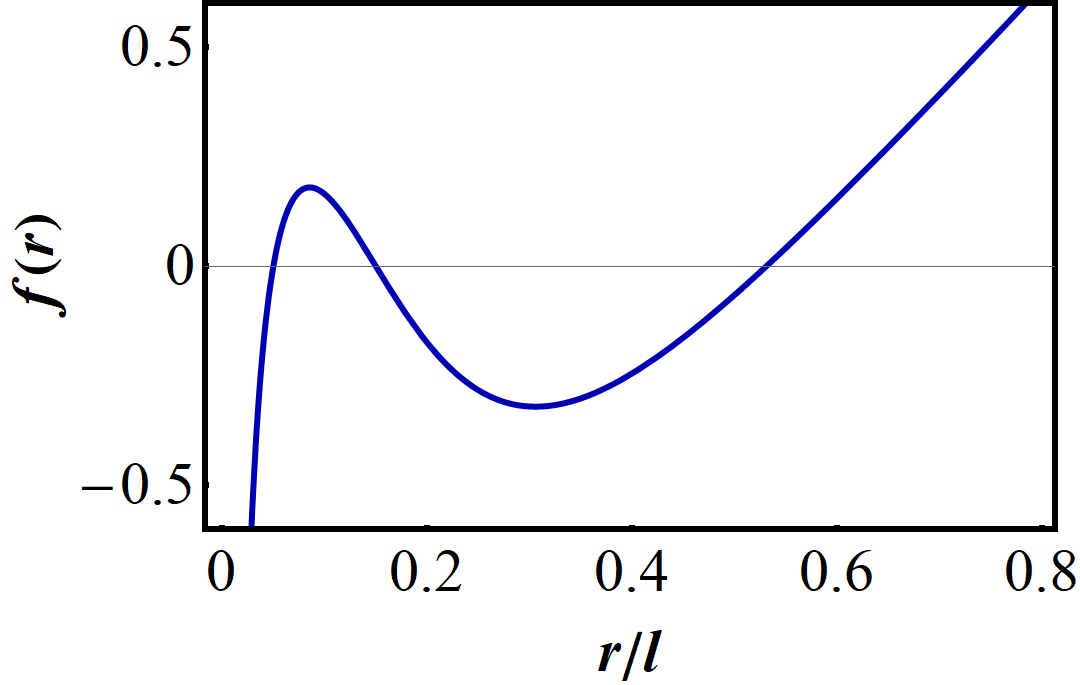}
        \caption{The blackening factor $f(r)$ }
        \label{fr3h}
    \end{subfigure}
    \caption{
    (a) The Penrose diagram for the three-horizon case. 
    The red area represents the minimum repetition unit (which can be interpreted as our observable universe). 
    I, II and III represent the three regions of the black hole interiors, respectively, where I corresponds to the region between the event horizon and the Cauchy horizon, II corresponds to the region between the Cauchy horizon and the third horizon, III corresponds to the region between the third horizon and the singularity. 
    $r_{\infty}$ stands for the spatial infinity and the wavy line at $r = 0$ indicates a spacelike singularity. 
    The gray grid area has no space-time region.
    (b) The blackening factor $f(r)$ for the Bardeen-AdS class black hole with $l = 5, \alpha = 1, s = 3, q = 1.3, M = 1.05 q^3/\alpha$, for which there are three zero points.
    }
   \label{BH3h}
\end{figure}

\begin{figure}[htbp]
    \centering
    \begin{subfigure}[b]{0.45\textwidth}
        \centering
		\includegraphics[scale=0.16]{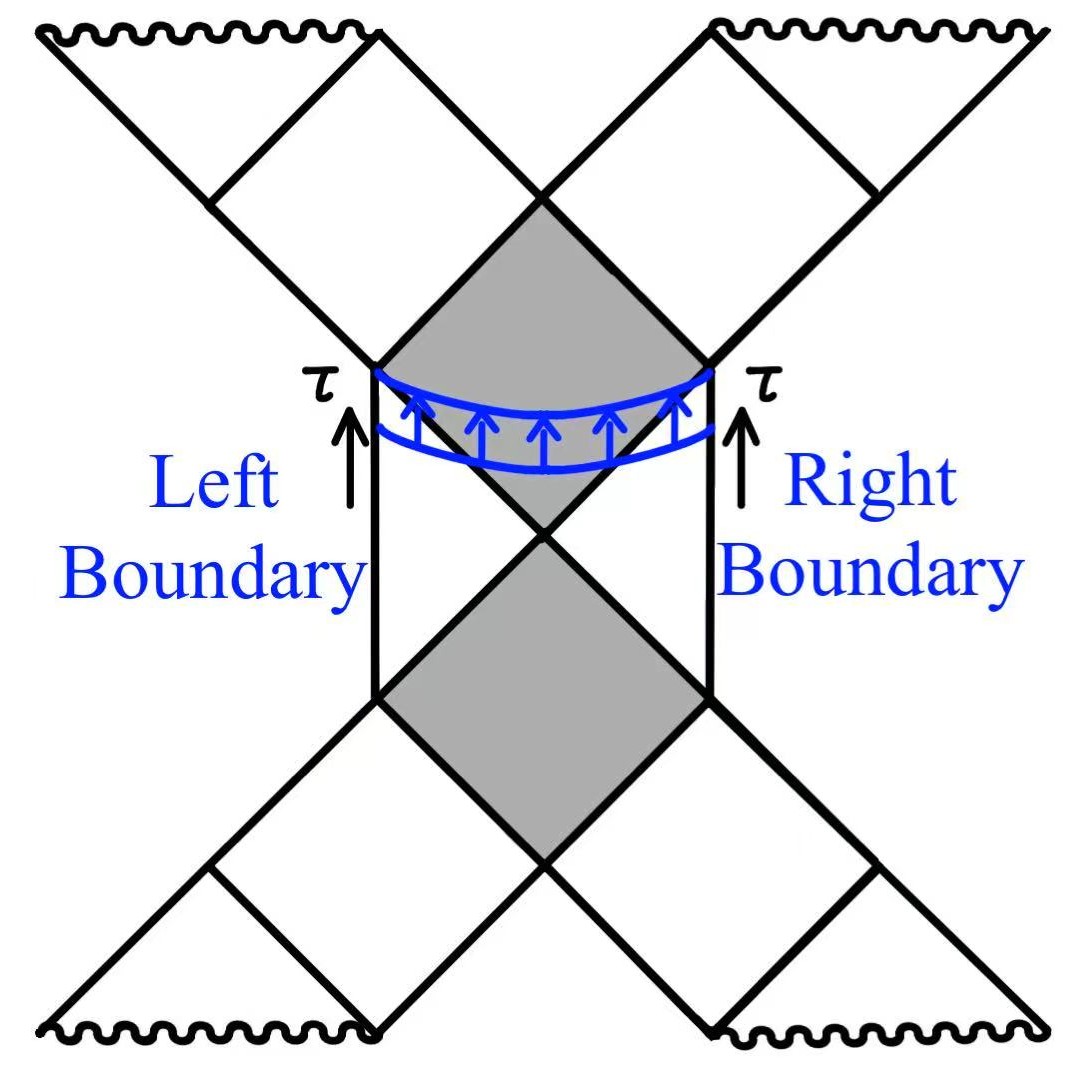}
        \caption{Common Region I}
        \label{connectionInI}
    \end{subfigure}
    \hspace{0.03\textwidth}
    \begin{subfigure}[b]{0.45\textwidth}
        \centering
		\includegraphics[scale=0.16]{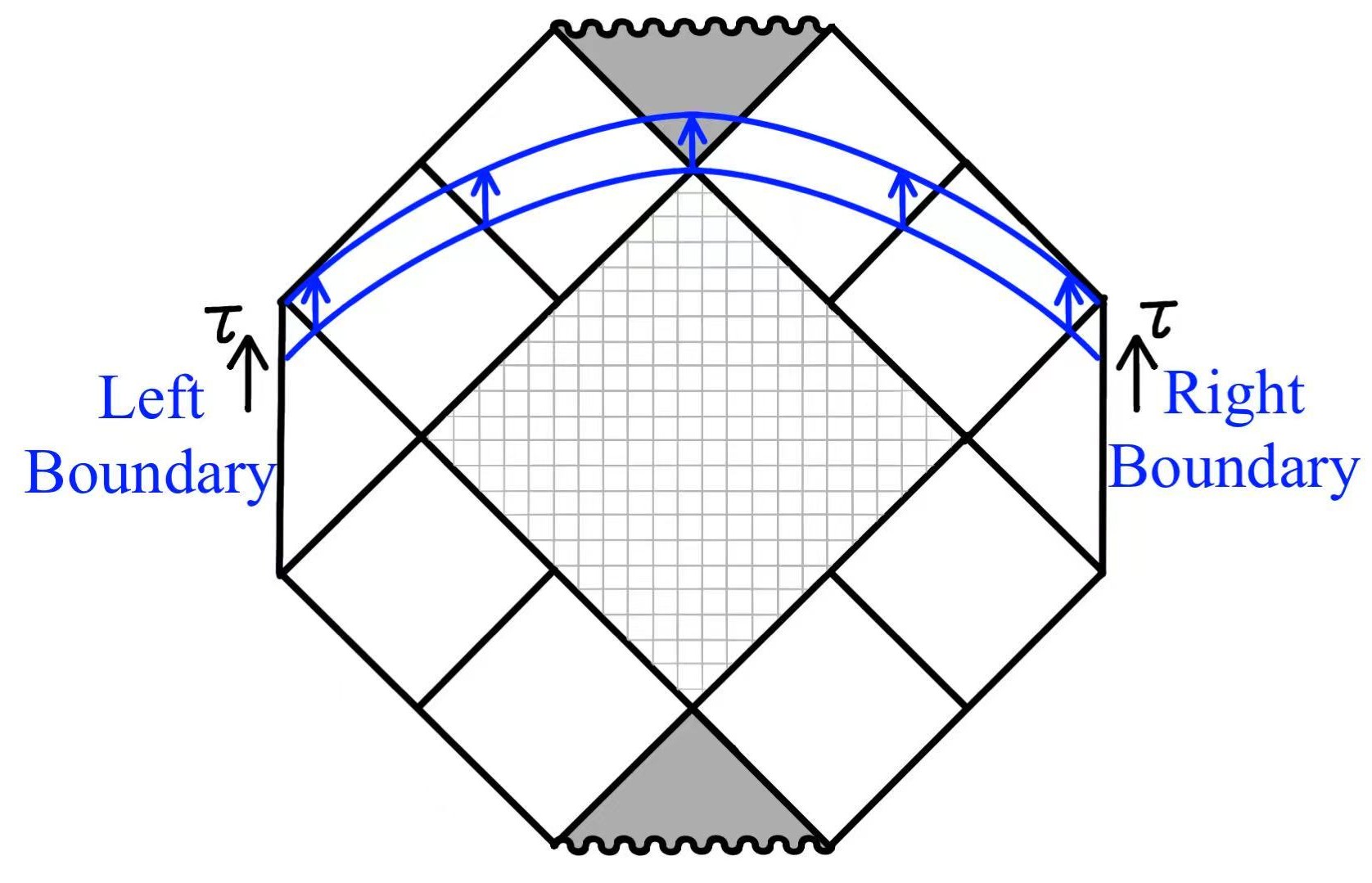}
        \caption{Common Region III}
        \label{connectionInIII}
    \end{subfigure}
    \caption[swx]{The Penrose like diagrams (without the maximum analytic continuation) for the AdS two-sided black hole in the three-horizon case. 
    The extremal surfaces are anchored at different left boundaries.
    (a) Two different universes have an identical Region I. 
    (b) Two different universes have an identical Region III. }
   \label{3Hconnection}
\end{figure}

In this subsection, we consider the three-horizon case in detail. 
As developed in Ref.~\cite{Yang:2021civ}, black holes satisfying the strong energy condition are limited to having at most one inner horizon. 
However, when the strong energy condition is violated---as in the theories of nonlinear electrodynamics~\cite{Gao:2021kvr}---black holes with multiple horizons are allowed. 
For the Bardeen-AdS class black hole, specific parameter choices can lead to the formation of three horizons. 
The corresponding Penrose diagram is shown in Fig.~\ref{PenroseDiagram3}, and the corresponding blackening factor $f(r)$ is shown in Fig~\ref{fr3h}. 
A complete version of the Penrose diagram for this model can be found in Ref.~\cite{Wu:2024gqi}. 
For a two-sided black hole with a fixed right spacetime boundary, Fig.~\ref{3Hconnection} illustrates that the left boundary is not unique. 
It means that there are two possible ways to splice the spacetime. The gray regions represent the overlapping spacetime domains where the Einstein-Rosen bridge grows continuously over time. 

\begin{figure}[htbp]
    \centering
    \begin{subfigure}[b]{0.45\textwidth}
        \centering
		\includegraphics[scale=0.16]{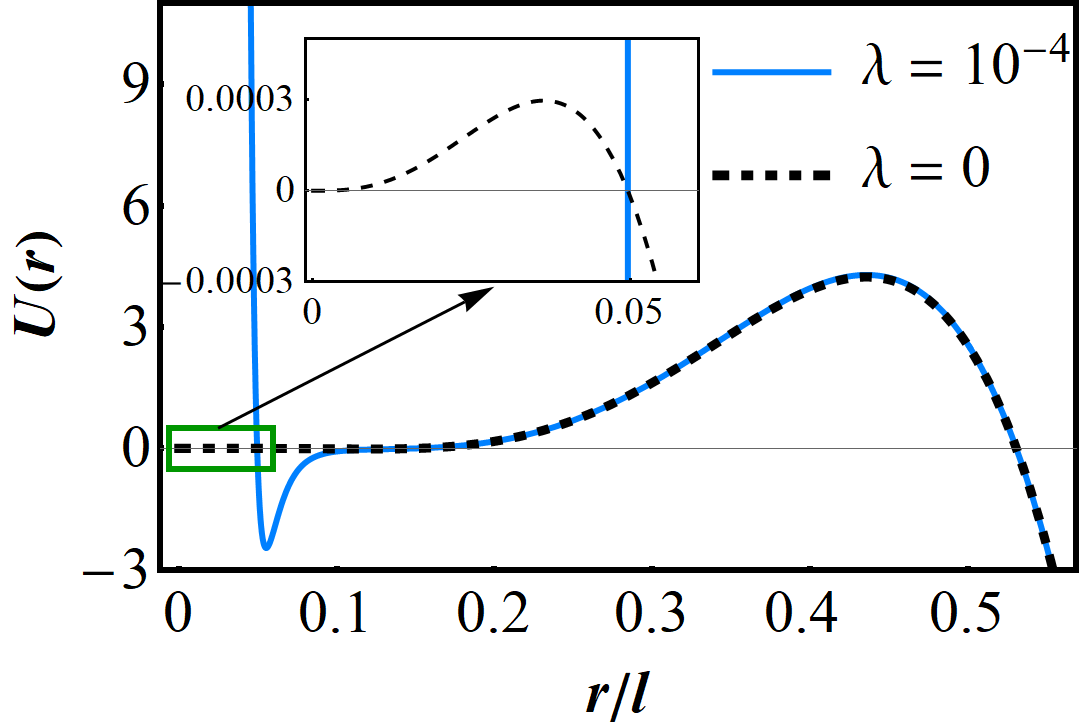}
        \caption{$a(r)= 1 + \lambda l^{4} C^{2}$}
        \label{Ur31}
    \end{subfigure}
    \hspace{0.05\textwidth}
    \begin{subfigure}[b]{0.45\textwidth}
        \centering
		\includegraphics[scale=0.16]{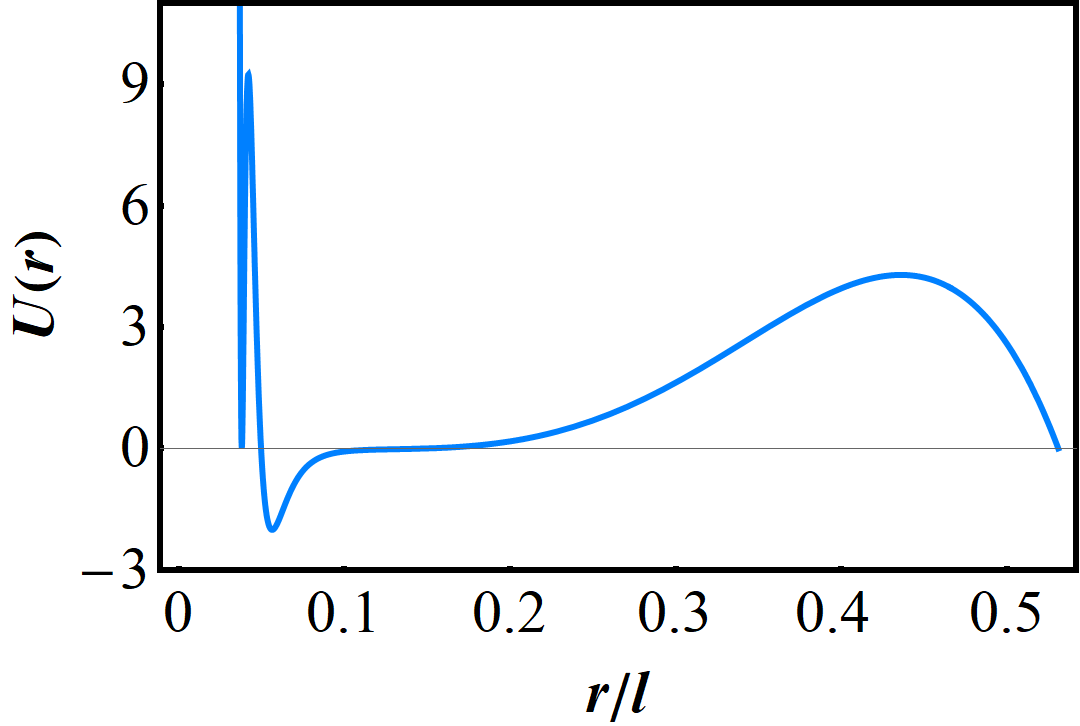}
        \caption{$a(r)= 1 + \lambda_{1} l^{4} C^{2}- \lambda_{2} l^{8} C^{4}$}
        \label{Ur32}
    \end{subfigure}
    \caption[swx]{The effective potentials with $l = 5, \alpha = 1, s = 3, M = 1.05\,q^3/\alpha, q = 1.3$.
    (a) The effective potentials have only one local maximum. When $\lambda=0$, the generalized volume complexity regresses to the CV proposal. 
    (b) $\lambda_{1} = 10^{-4}, \lambda_{2}=1.3\times 10^{-11}$. The effective potential has two local maxima.
    }
   \label{Ur3}
\end{figure}

In terms of the holographic complexity, we still use Eqs.~\eqref{ar1} and \eqref{ar2} to construct the gravitational observables.
The corresponding effective potentials are illustrated in Fig.~\ref{Ur3}. 
In the case shown in Fig~\ref{Ur31}, the local maximum appears only outside the Cauchy horizon, i.e., $r_{h_{2}} < r_{f} < r_{h_{1}}$. 
Notably, when $\lambda=0$, the holographic complexity regresses to the CV conjecture. 
In this case, the peak of the effective potential lies outside the Cauchy horizon, while the potential in the deeper region remains significantly lower. 
As a result, small perturbations originating from this region cannot overcome the barrier formed by the higher external potential. 
This implies that the extremal surface associated with the CV conjecture can only probe the region outside the Cauchy horizon, rendering the description incomplete. 
In Fig.~\ref{Ur32} case, the local maxima appear in regions $0<r_{f}<r_{h_{3}}$ and $r_{h_{2}}<r_{f}<r_{h_{1}}$. 
For such an effective potential, we can analyze the time evolution of the generalized volume complexity in Regions I and III, simultaneously. 
In the following discussion, we will focus on considering the case shown in Fig.~\ref{Ur32}. 
It should be pointed out that although Fig.~\ref{Ur32} resembles Figs.~\ref{Ur12} and \ref{Ur22}, there is no phase transition here. 
Instead, there are two independent surfaces anchored at different left boundary and without competing relationships. 

\begin{figure}[htbp]
    \centering
    \begin{subfigure}[b]{0.45\textwidth}
        \centering
        \includegraphics[scale=0.16]{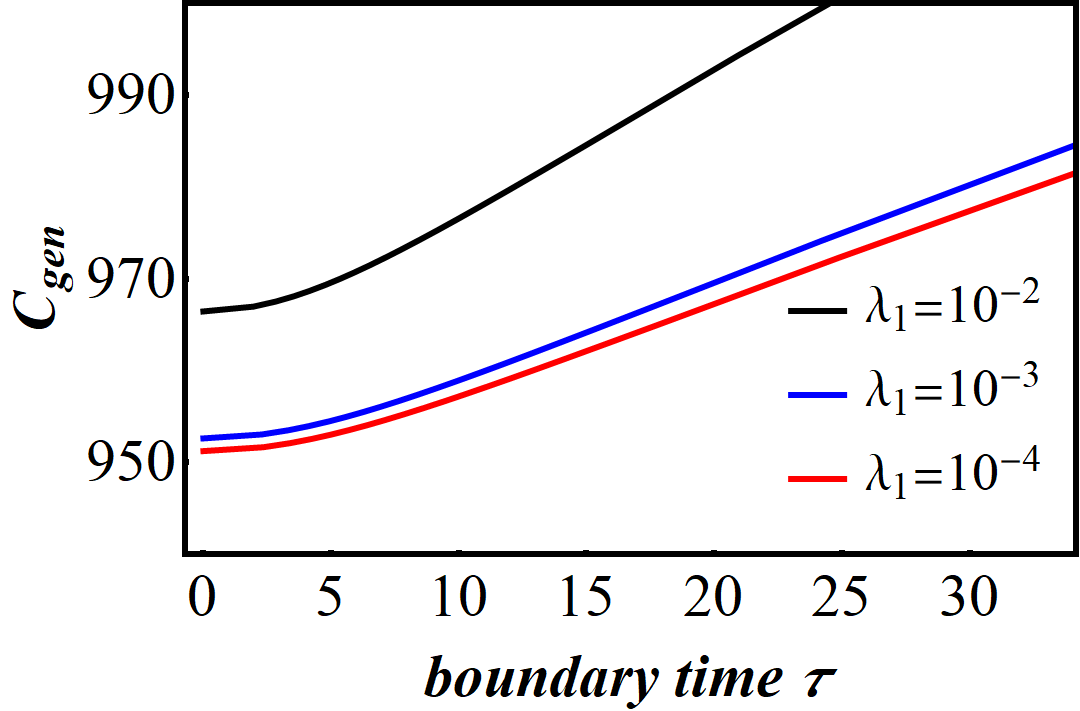}
        \caption{$a(r)= 1 + \lambda l^{4} C^{2}$}
    \end{subfigure}
    \hspace{0.05\textwidth}
    \begin{subfigure}[b]{0.45\textwidth}
        \centering
		\includegraphics[scale=0.15]{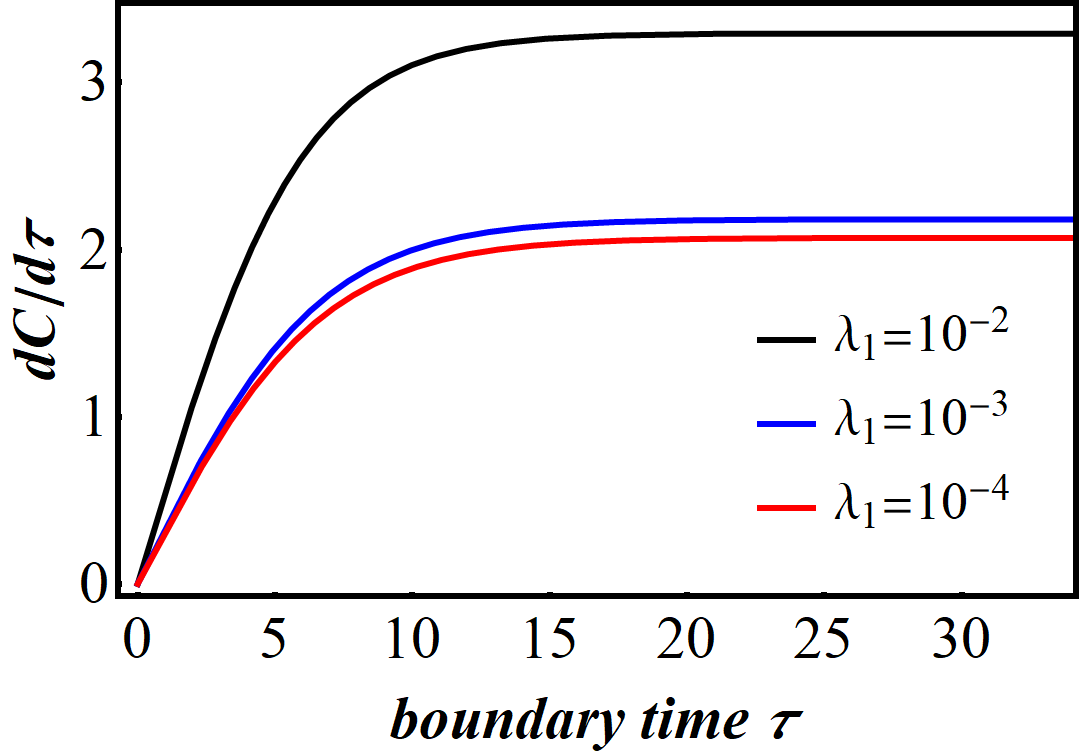}    
        \caption{$a(r)= 1 + \lambda_{1} l^{4} C^{2}- \lambda_{2} l^{8} C^{4}$}
    \end{subfigure}
    \caption[swx]{Time evolution of the generalized volume complexity under different values of $\lambda_{1}$. }
    \label{dcdt31}
\end{figure}

When $r_{h_{2}}<r_{f}<r_{h_{1}}$, the extremal surfaces are anchored at two boundary times that are relatively close together as shown in Fig.~\ref{connectionInI}. 
In this case, the complexity probes Region I, and we can obtain the complexity growth as shown in Fig.~\ref{dcdt31}. 
An excessively large $\lambda_{2}$ will cause the local maximum near the singularity to disappear, which is undesirable. 
A sufficiently small $\lambda_{2}$ has little effect on the local maximum on the right, 
so we can only discuss the influence of $\lambda_{1}$ on the complexity growth rate.
Due to the Cauchy horizon, the local maximum of the effective potential always exists, and $\lambda_{1}$ is desirable in the entire parameter space. 
At late times, the complexity growth rate exhibits a positive correlation with increasing $\lambda_{1}$. 

\begin{figure}[htbp]
    \centering
    \begin{subfigure}[b]{0.45\textwidth}
        \centering
        \includegraphics[scale=0.15]{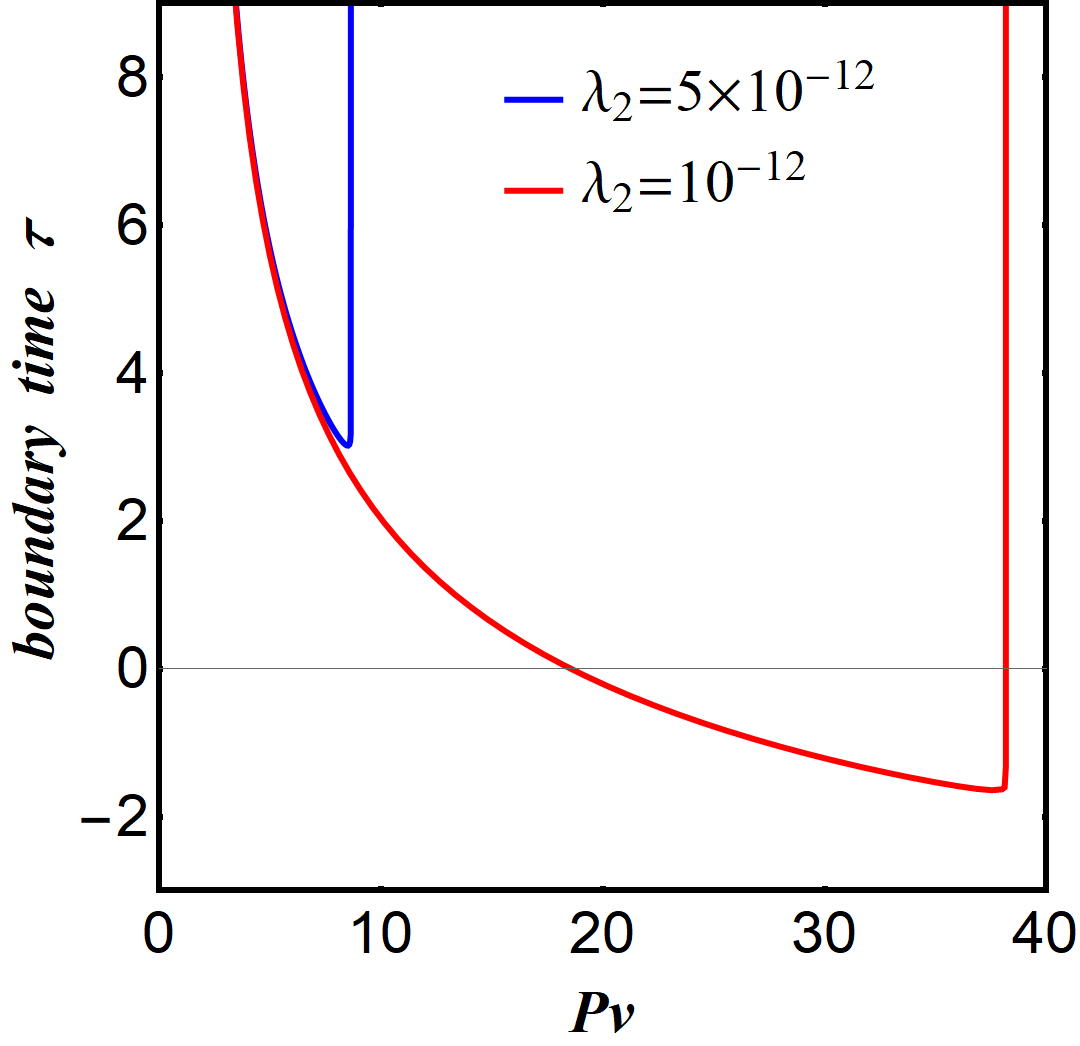}
        \caption[swx]{$P_{v}\sim \tau$}
        \label{pvtpm}
    \end{subfigure}
    \begin{subfigure}[b]{0.45\textwidth}
        \centering
        \includegraphics[scale=0.143]{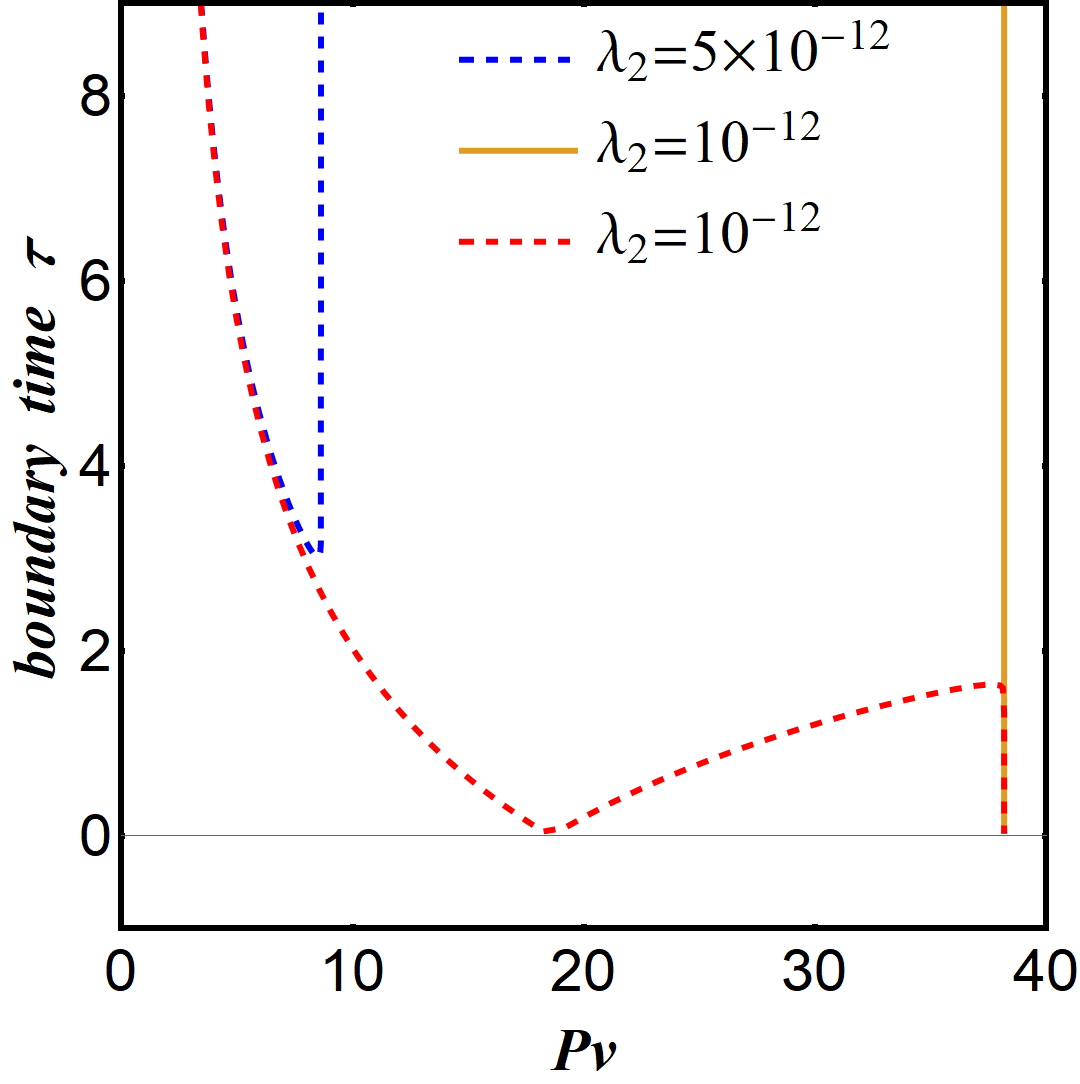}
        \caption[swx]{Suitable Surface}
        \label{RealSurface}
    \end{subfigure}
    \caption[swx]{The relation between the boundary time $\tau$ and the conserved momentum $P_{v}$ with $l = 5, \alpha = 1, s = 3, M = 1.05\,q^3/\alpha, q = 1.3, \lambda_{1}=10^{-4}.$
    }
   \label{pvt3322}
\end{figure}

When $0<r_{f}<r_{h_{3}}$, the extremal surfaces are anchored at two boundaries that are relatively far apart as shown in Fig.~\ref{connectionInIII}, and the complexity probes the Region III. 
In this case, the extremal surfaces may not evolve from $\tau=0$ as shown in Fig.~\ref{pvtpm}, which is inappropriate. 
It should be noted that Fig.~\ref{pvtpm} is not a complete $P_{v}-\tau$ diagram, but only includes the part where $0<r<r_{h_{3}}$. 
The blue branch indicates that the extremal surface evolves starting at a finite boundary time, rather than at zero. 
This implies that the complexity will evolve starting at a non-zero boundary time, which is flawed. 
The red branch indicates that the extremal surfaces evolve starting at a negative boundary time.
In fact, due to time reversal symmetry, the negative time portion can be viewed as two surfaces evolving positively from zero to a finite value, and extending into the white hole~\cite{Myers:report}. 
We do not consider these surfaces appropriate for defining the complexity. 
Since the dipping branches have been proven to be non-physical~\cite{Belin:2022xmt, Omidi:2022whq, Wang:2023eep}, we will not discuss them further in this work. 
Therefore, the only case we need to consider is the extremal surface that evolves starting at $\tau = 0$ and approaches the singularity, as shown in Fig.~\ref{RealSurface}. 
\begin{figure}[htbp]
    \centering
    \begin{subfigure}[b]{0.45\textwidth}
        \centering
        \includegraphics[scale=0.16]{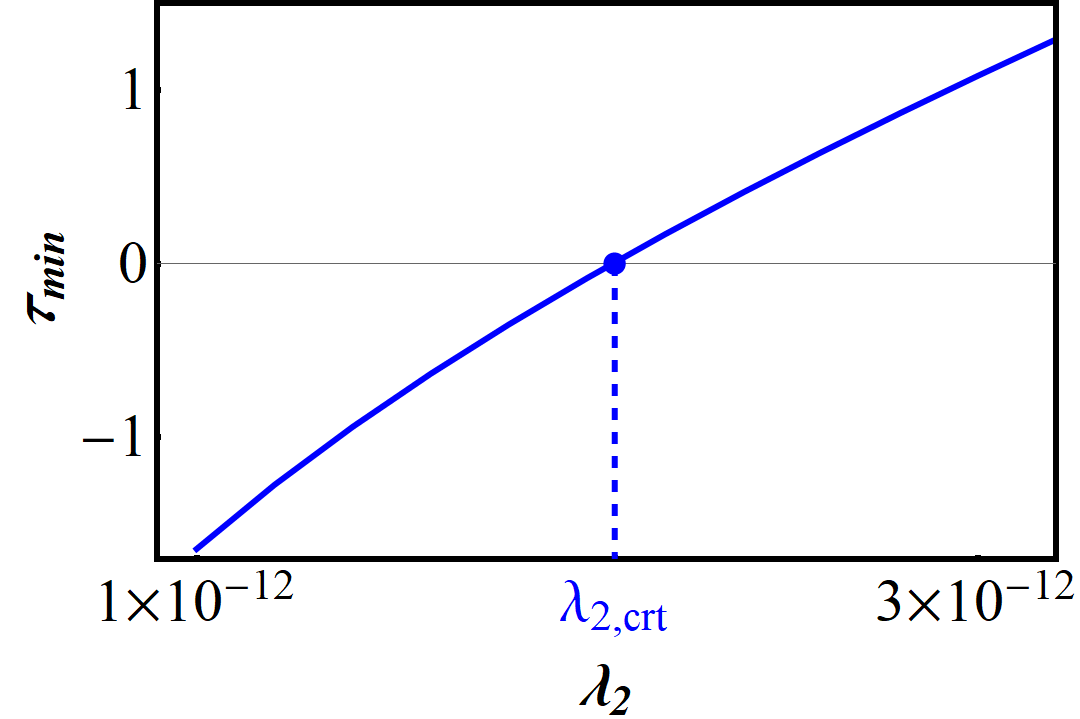}
        \caption[swx]{The minimum value of $\tau$}
        \label{tmin}
    \end{subfigure}
     \hspace{0.03\textwidth}
    \begin{subfigure}[b]{0.45\textwidth}
        \centering
        \includegraphics[scale=0.17]{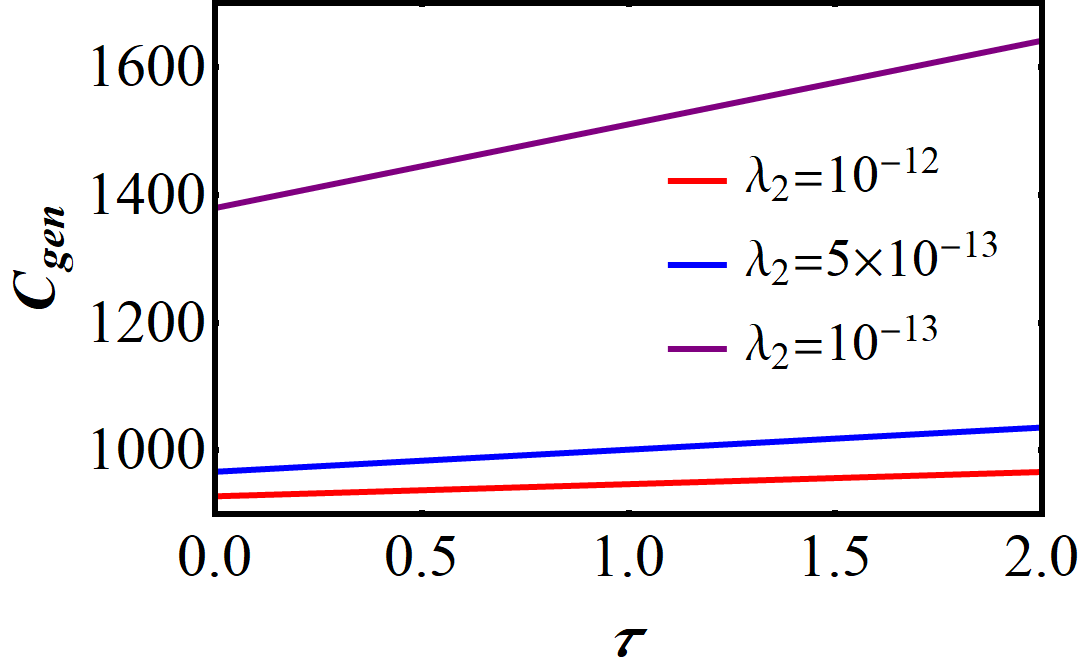}
        \caption[swx]{Time evolution of $\mathcal{C}_{\text{gen}}$}
        \label{cgent32}
    \end{subfigure}
    \caption[swx]{The minimum value of $\tau$ and the time evolution of the generalized volume complexity with $l = 5, \alpha = 1, s = 3, M = 1.05\,q^3/\alpha, q = 1.3, \lambda_{1}=10^{-4}.$}
   \label{Cgen3H2P}
\end{figure}

\begin{figure}[htbp]
    \centering
    \begin{subfigure}[b]{0.45\textwidth}
        \centering
        \includegraphics[scale=0.16]{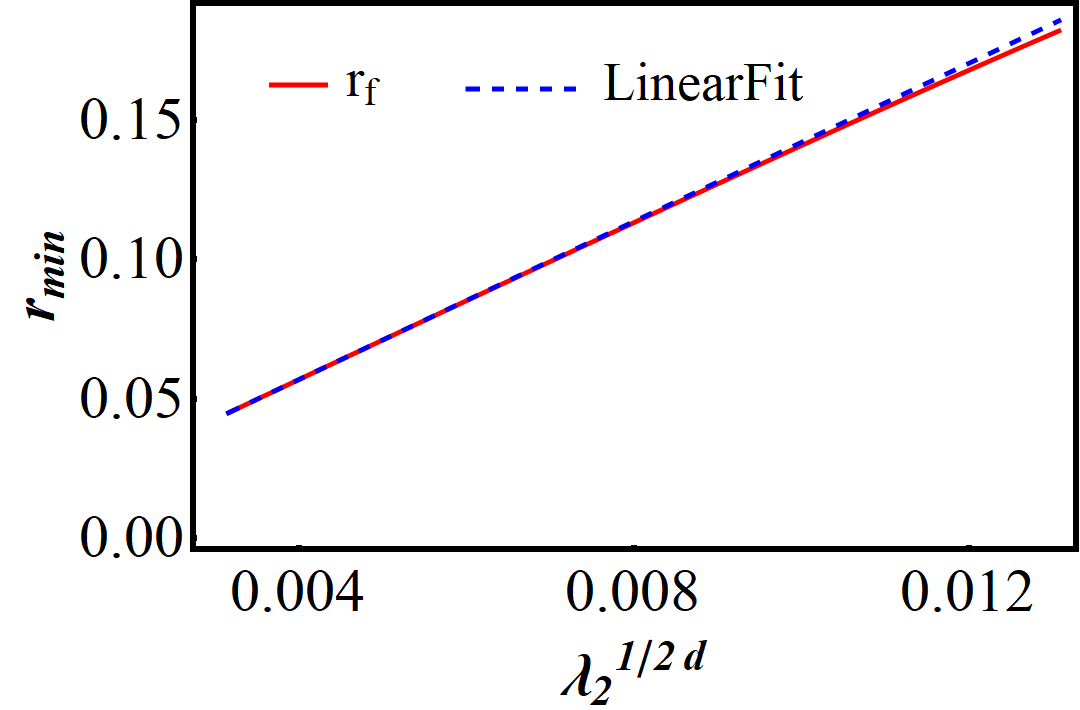}
        \caption[swx]{Linear scaling of $r_{f}$ for small $\lambda_2$}
        \label{rfandfix}
    \end{subfigure}
    \begin{subfigure}[b]{0.45\textwidth}
        \centering
        \includegraphics[scale=0.16]{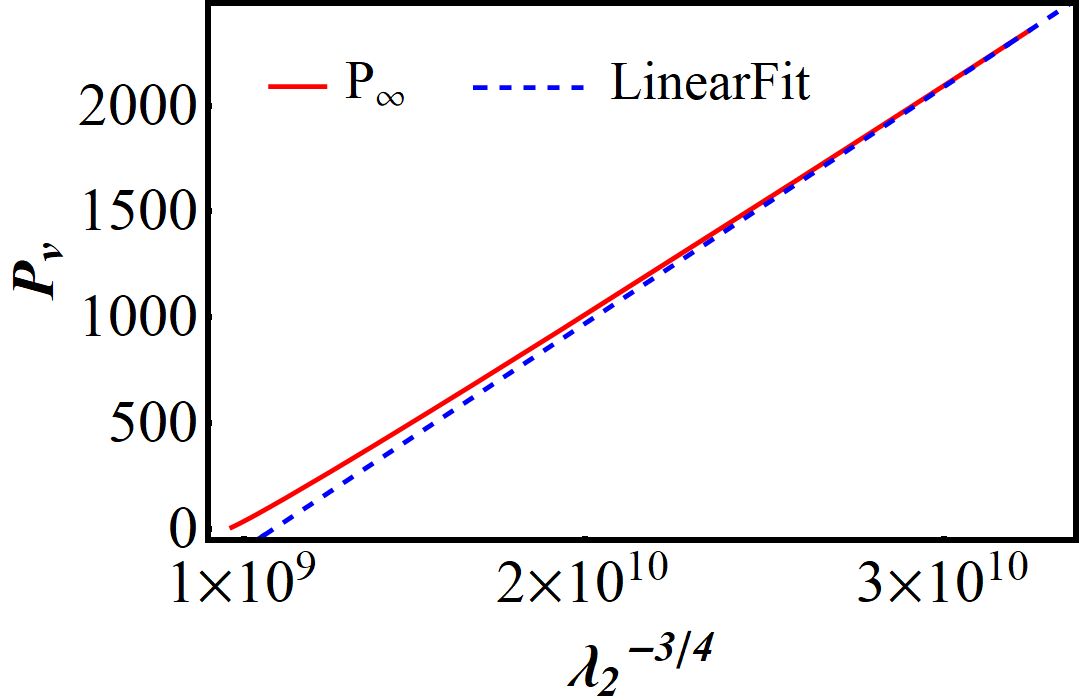}
        \caption[swx]{Linear scaling of $P_{\infty}$ for small $\lambda_2$}
        \label{Pvlambda2}
    \end{subfigure}
    \caption[swx]{The linear behavior in the small $\lambda_2$ regime with $l = 5, \alpha = 1, s = 3, M = 1.05\,q^3/\alpha, q = 1.3, \lambda_{1}=10^{-4}.$
    (a) The red curve shows the numerically computed $r_{f}$ as a function of $\lambda_2^{1/2d}$, while the blue dashed line represents a linear fit in the $\lambda_2 \to 0 $ limit. 
    (b) The red curve shows the numerically computed $P_{\infty}$ as a function of $\lambda_2^{-3/4}$, while the blue dashed line represents a linear fit in the $\lambda_2 \to 0 $ limit. 
    }
   \label{probes}
\end{figure}

The constraint that the extremal surface originates from $\tau=0$ imposes an upper bound on $\lambda_{2 }$, i.e., $\lambda_{2} \leqslant \lambda_{2,\text{crt}}$. 
This upper bound can be obtained by calculating the minimum value of the boundary time (without considering the time reversal symmetry) as shown in Fig.~\ref{tmin}. 
We can then easily obtain the time evolution of $\mathcal{C}_{\text{gen}}$, as shown in Fig.~\ref{cgent32}. 
Fixing $\lambda_{1}$, the $r_{f}$ decreases as $\lambda_{2}$ decreases. 
When $\lambda_{2} \to 0$, the extremal surface approaches the singularity. 
Reference~\cite{Jorstad:2023kmq} derived the analytical solution of $r_{f}$ for the planar AdS black hole. 
While an analytical expression for our model remains elusive, the relation $r_{f} \propto \lambda_{2}^{1/2d}$ is still valid, as demonstrated in Fig.~\ref{rfandfix}. 
As the surface approaches the singularity, the complexity growth rate exhibits a scaling relation $\frac{d\mathcal{C}_{\text{gen}}}{d\tau}\Big|_{\tau \to \infty} \varpropto \lambda_{2}^{-3/4}$, as shown in Fig.~\ref{Pvlambda2}. 

In conclusion, we have demonstrated the unique advantages of the generalized volume complexity in exploring the internal structure of multi-horizon black holes, 
using a three-horizon black hole as an example. 
In Sec.~\ref{sec6}, we will extend this result to a more general case.

\section{Probing the singularity or the Cauchy horizon with CMC slices}\label{sec5}
In this section, we use the ``complexity equals anything" proposal with codimension-zero observables, to analyze the choice of space-time boundaries and probe the singularity. 
Following the methodology of Ref.~\cite{Jorstad:2023kmq}, we focus on the future CMC slice $\Sigma_{+}$ and evaluate the observables with $F_{1,-}=G_{1}=0$. 
Then, the expression of the complexity can be given by
\begin{equation}
    \mathcal{C}_{\text{gen}}^{+}=\frac{1}{G_{\text{N}}l}\int_{\Sigma_{+}(\alpha_{+},\alpha_{B})}\,d^{d}\sigma\,\sqrt{h}\,a(r). 
\label{CCMC}
\end{equation}
As $\tau\to\infty$, we can obtain the growh rate of the complexity at late times, i.e.,
\begin{equation}
    \lim_{\tau\to\infty}\frac{d}{d\tau} \mathcal{C}_{\text{gen}}^{+}=\frac{V_{d-1}}{G_{\text{N}}l}\,r_{f}^{d-1}\,a(r_{f})\,\sqrt{-f(r_{f})}.
\label{dcplusinfty}
\end{equation}

\subsection{Extremal surfaces}
\begin{figure}[htbp]
    \centering
    \begin{subfigure}[b]{0.45\textwidth}
        \centering
        \includegraphics[scale=0.16]{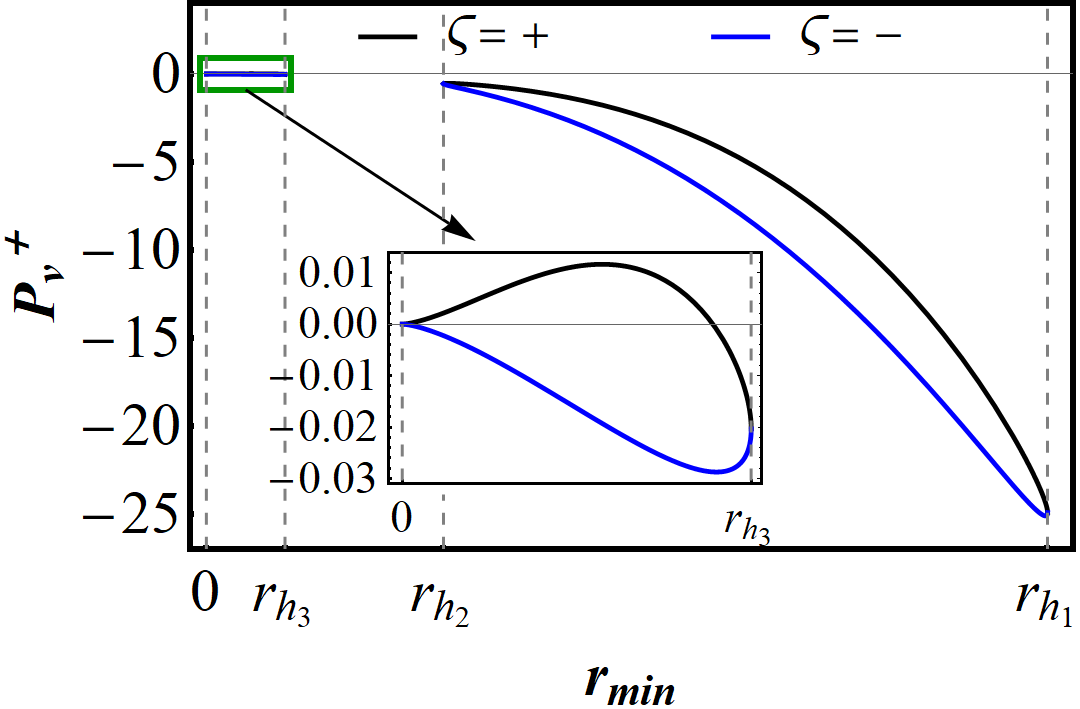}
        \caption[swx]{$\alpha_{B}=20$}
        \label{FigPvrma}
    \end{subfigure}
    \begin{subfigure}[b]{0.45\textwidth}
        \centering
        \includegraphics[scale=0.155]{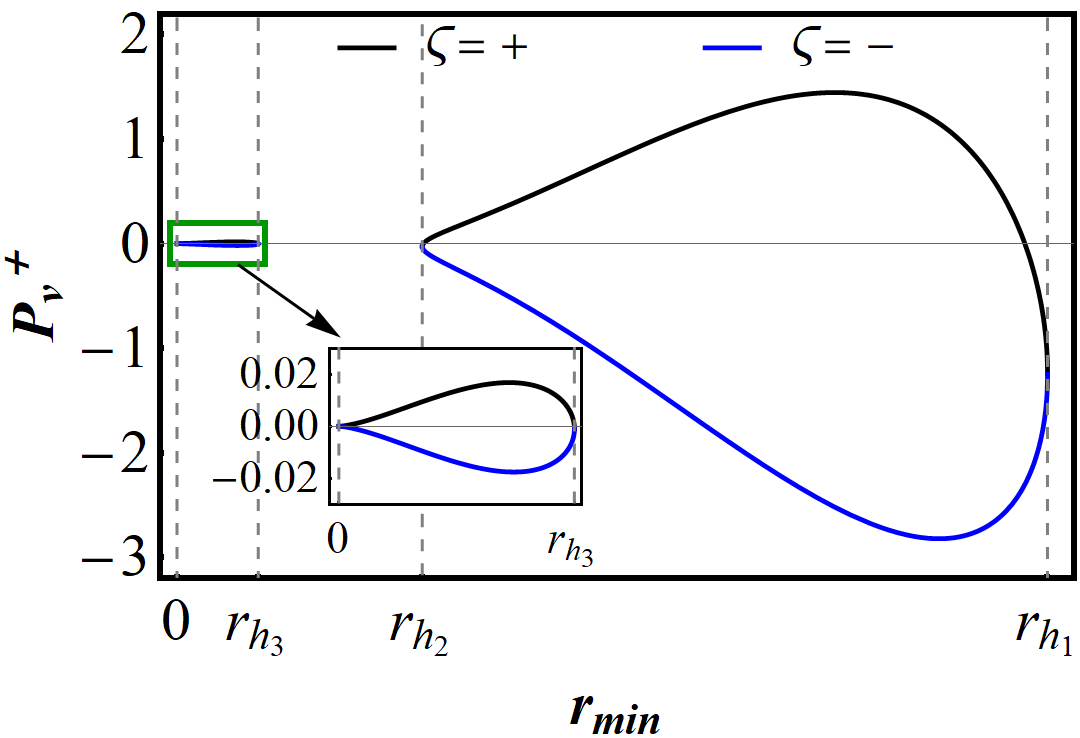}
        \caption[swx]{$\alpha_{B}=1$}
        \label{FigPvrmb}
    \end{subfigure}
    \caption[swx]{The conserved momentum with $l = 5, \alpha = 1, s = 3, M = 1.05\,q^3/\alpha, q = 1.3, \alpha_{+}=1.$}
    \label{FigPvrm}
\end{figure}

\begin{figure}[htbp]
    \centering
    \begin{subfigure}[b]{0.45\textwidth}
        \centering
        \includegraphics[scale=0.2]{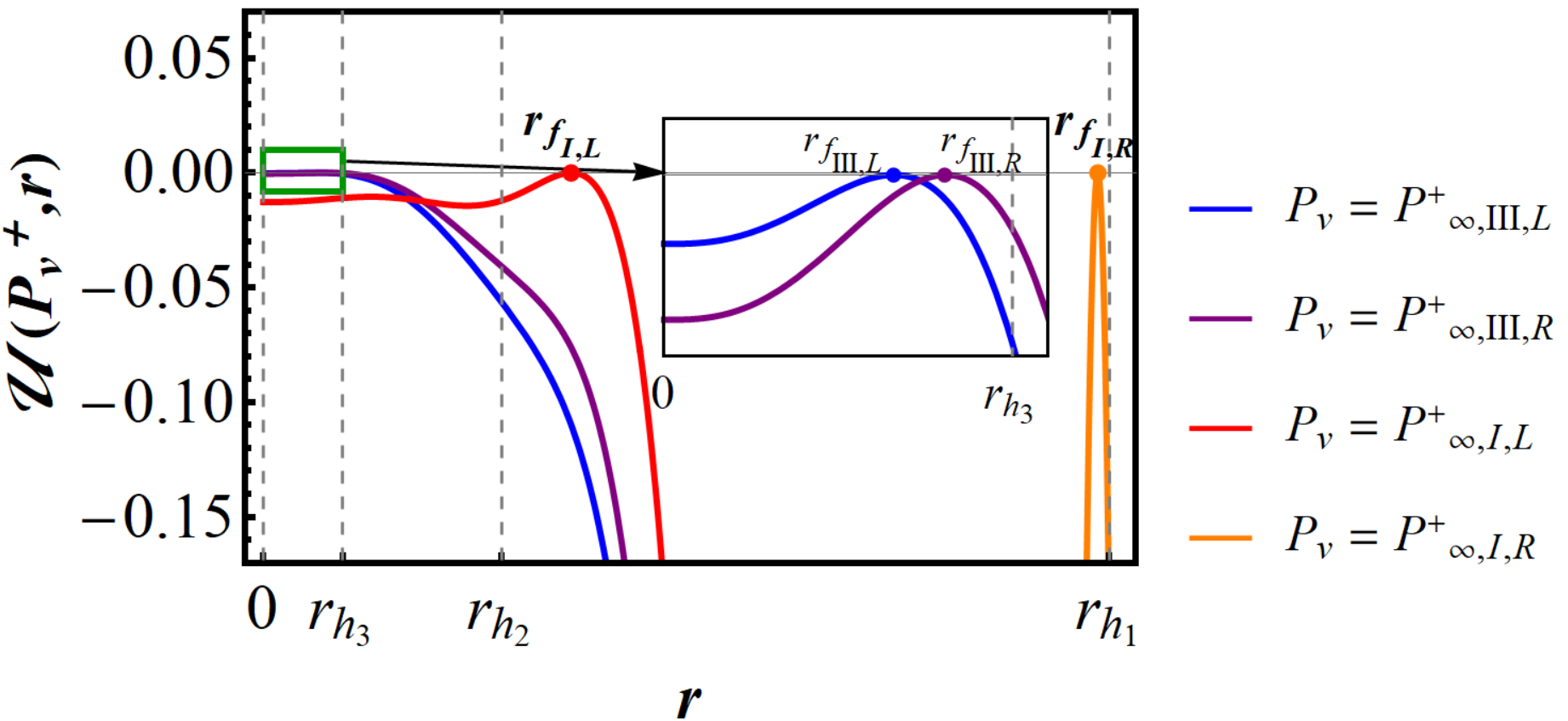}
        \caption[swx]{Effective potentials with $\alpha_{B}=8$}
        \label{Urrf8}
    \end{subfigure}
    \hspace{0.05\textwidth}
    \begin{subfigure}[b]{0.45\textwidth}
        \centering
        \includegraphics[scale=0.12]{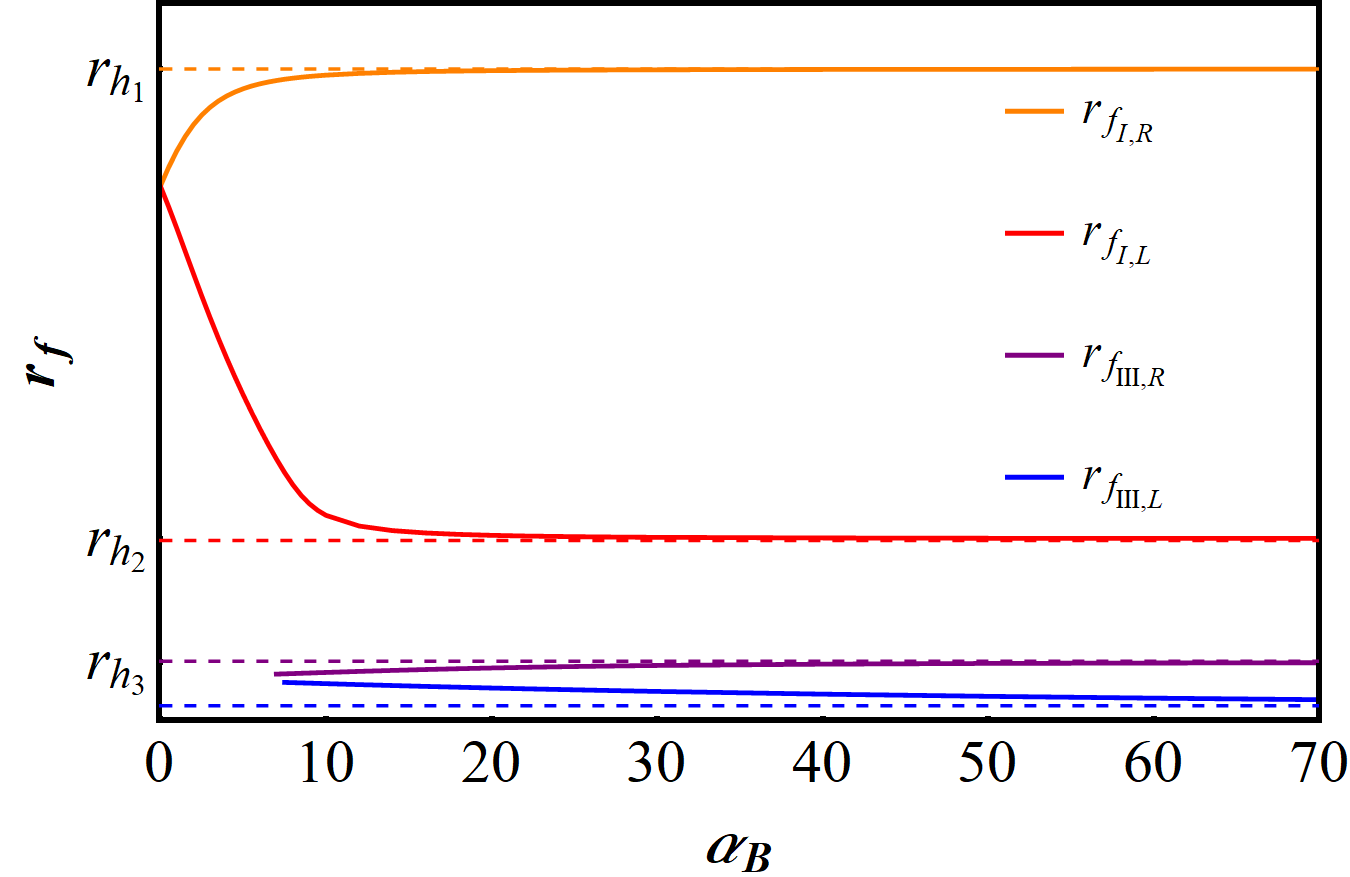}
        \caption[swx]{The position of $r_{f}$ depends on $\alpha_{B}$}
        \label{Urrfab}
    \end{subfigure}
    \caption[swx]{The effective potential with $l = 5, \alpha = 1, s = 3, M = 1.05\,q^3/\alpha, q = 1.3, \alpha_{+}=1, \alpha_{B}=8.$
    (a) When $\alpha_{B}$ is large enough (in our example, $\alpha_{B}=8$), there are four extremal surfaces that can evolve to late times. 
    (b) 
    As $\alpha_{B}$ increases, $r_{f}$ gradually approach the horizons or the spacelike singularity. 
    }
    \label{Urrf}
\end{figure}

We still begin our discussion by considering the Bardeen-AdS class black hole as a concrete example. 
The situations of one horizon and two horizons differ from Ref.~\cite{Jorstad:2023kmq} only in the blackening factor, 
so we only consider the three-horizon case here. 
The extremization of the surfaces has been discussed in Sec.~\ref{sec3.2}, we can therefore control the distance between the CMC slice and the singularity (or the Cauchy horizon) by varying $\alpha_{\pm}$ or $\alpha_{B}$. 
In this paper, we will fix $\alpha_{+}$ and vary $\alpha_{B}$. Therefore, Eq.~\eqref{br} can be written as
\begin{equation}
    b(r)=\frac{\alpha_{B}}{3l\alpha_{+}}r^{3}.
\end{equation}
With $\alpha_{B}$ fixed, we numerically determine the relationship between $P_{v}^{+}$ and $r_{\text{min}}$, as illustrated in Fig.~\ref{FigPvrm}. 
Similar to the codimension-one case, for a given $P_{v}$, only the largest $r_{\text{min}}$ can be taken. 
Therefore, the case shown in Fig.~\ref{FigPvrmb} cannot probe Region III, whereas the case shown in Fig.~\ref{FigPvrma} can. 
 {This requires a lower bound on $\alpha_{B}$, i.e., $\alpha_{B}>\alpha_{Bmin}$, which is fixed by the condition that $P_{v}^{+}$ develops two equally high local maxima.
In other words, only when the extrinsic curvature of the CMC slice is large enough can Region III be probed. }
When $\alpha_{B} \to \infty$ , the future CMC slices approach null surfaces. 
At late times, using Eq.~\eqref{tauinfty}, we can compute the effective potential $\mathcal{U}(P_{v},r_{f})$ for a given $\alpha_{B}$ as shown in Fig.~\ref{Urrf8}.
It is important to note that for each region within the spacetime, the outer surfaces correspond to $ \zeta = - $, which are anchored at $\tau \to -\infty$. 
In this section, we only discuss the surfaces with positive boundary times, i.e., two surfaces determined by $P^{+}_{\infty, III, L}$ and $P^{+}_{\infty, I, L}$. 

\subsection{Singularity and Cauchy horizon probes}\label{sec5.2}
When $\alpha_{B} \to \infty$, the extremal surfaces approach the singularity or the Cauchy horizon. 
For the singularity branch, we can obtain
\begin{equation}
    r_{f}=(2l/3)^{2/3}(2M-2q^{3}/\alpha)^{1/3} (\alpha_{+}/\alpha_{B})^{2/3} .
\label{rf}
\end{equation}
It should be pointed out that Eq.~\eqref{rf} is only valid for $M>q^{3}/\alpha$.
By substituting Eq.~\eqref{rf} back to Eq.~\eqref{dcplusinfty} and choosing different expressions of $a(r)$, we can obtain
\begin{align}
    a(r)=1\,\,\,:\,\,\,\lim_{\tau\to\infty}&\frac{d}{d\tau} \mathcal{C}_{\text{gen}}^{+}=\frac{3 V_{d-1}}{G_{\text{N}}}\left(M-q^{3}/\alpha\right)(\alpha_{+}/\alpha_{B})\to 0, \label{dccmc1}\\
    a(r)=|lK|\,\,\,:\,\,\,\lim_{\tau\to\infty}&\frac{d}{d\tau} \mathcal{C}_{\text{gen}}^{+}=\frac{3 V_{d-1}}{G_{\text{N}}}\left(M-q^{3}/\alpha\right)\to \text{Const.},\label{dccmc2}\\
    a(r)=|l^4C^2|\,\,\,:\,\,\,\lim_{\tau\to\infty}&\frac{d}{d\tau} \mathcal{C}_{\text{gen}}^{+}\backsim (\alpha_{B}/\alpha_{+})^{3}\to \infty,\label{dccmc3}
\end{align}
where the trace of the extrinsic curvature $K$ is given in Eq.~\eqref{ExCur}, 
and the square of the Weyl tensor $C^{2}$ is given in Eq.~\eqref{WeylTensor2}. 
In fact, this conclusion also applies to the case of a single horizon. 
For the Cauchy horizon branch, when $\alpha_{B}\to\infty$, $r_{f}\to r_{h_{2}}$, that is, $f(r_{f})\to 0$. 
Based on these two limits, it is easy to derive 
\begin{align}
    \frac{4 \alpha_{B}^{2}}{l^2 \alpha_{+}^2}f(r_{f}) + f'^{2}(r_{f})\simeq 0,\label{rfh21}\\
    r_{f}\simeq r_{h_{2}}-\frac{l^{2} \alpha_{+}^{2}}{4\alpha_{B}^{2}}f'(r_{h_{2}})\label{rfh22},
\end{align}
where Eq.~\eqref{rfh22} is derived by performing a first-order expansion of Eq.~\eqref{rfh21}.
For non-extremal black holes, the necessary condition for Eq.~\eqref{rfh21} to have real solutions is $f(r_{f})<0$, and thus, the horizon that $r_{f}$ approaches must be $r_{h_{2}}$. 
By substituting Eqs.~\eqref{rfh21} and \eqref{rfh22} into Eq.~\eqref{dcplusinfty}, we can still determine the complexity growth rate at late times: 
\begin{equation}
    \lim_{\tau\to\infty}\frac{d}{d\tau} \mathcal{C}_{\text{gen}}^{+}=\frac{V_{d-1}}{2G_{\text{N}}}\,|f'(r_{h_{2}})| r_{h_{2}}^{2}\frac{\alpha_{+}}{\alpha_{B}}a(r_{f}).
\label{dcrh2}
\end{equation}
Due to the nonanalytic horizons, it is challenging to derive an explicit analytical expression for Eq.~\eqref{dcrh2}. 
However, we can still get three limits similar to Eqs.~\eqref{dccmc1}, \eqref{dccmc2} and \eqref{dccmc3}, i.e., 
\begin{align}
    a(r)=1\,\,\,:\,\,\,\lim_{\tau\to\infty}&\frac{d}{d\tau } \mathcal{C}_{\text{gen}}^{+}\backsim(\alpha_{+}/\alpha_{B})\to 0,\label{dccmc21}\\
    a(r)=|lK|\,\,\,:\,\,\,\lim_{\tau\to\infty}&\frac{d}{d\tau} \mathcal{C}_{\text{gen}}^{+} \to \text{Const.},\label{dccmc22}\\
    a(r)=|l^4C^2|\,\,\,:\,\,\,\lim_{\tau\to\infty}&\frac{d}{d\tau} \mathcal{C}_{\text{gen}}^{+}\backsim (\alpha_{+}/\alpha_{B})^{3}\to 0,\label{dccmc23}
\end{align}

To summarize, as the extremal surfaces approach null surfaces, the volume naturally vanishes.
Consequently, the complexity growth rate at late times, derived from Eq.~\eqref{dccmc1} or \eqref{dccmc21}, tends to zero. 
Due to the divergence of the extrinsic curvature, the complexity growth rate derived from Eq.~\eqref{dccmc2} or \eqref{dccmc22} remains finite at late times. 
The difference between Eq.~\eqref{dccmc3} and Eq.~\eqref{dccmc23} reflects the geometric properties of the curvature singularity. 
In general, we can choose 
\begin{equation}
    a(r)=\lambda_{1} + \lambda_{2}|lK| + \lambda_{3}|l^{4}C^{2}|, 
\end{equation}
where $\lambda_{i}$ are dimensionless parameters. 
When $\lambda_{1}$, $\lambda_{2}$, and $\lambda_{3}$ are all finite and non-zero, that is, when all three contributions are considered simultaneously, 
the complexity growth rate at late times reduces to
\begin{equation}
    \left\{
    \begin{aligned}
        \lim_{\tau\to\infty}&\frac{d}{d\tau } \mathcal{C}_{\text{gen}}^{+}\to \text{Const.},\,\,\,\,\,\,\,\,r_{f}\to r_{h_{2}},\\
        \lim_{\tau\to\infty}&\frac{d}{d\tau } \mathcal{C}_{\text{gen}}^{+}\to \infty,\,\,\,\,\,\,\,\,\,\,\,\,\,\,\,\,\,\,r_{f}\to 0.
    \end{aligned}
    \right.
    \label{CgenWhole}
\end{equation}
Eq.~\eqref{CgenWhole} shows that distinguishing the two regions is feasible using different constructions of the scalar function $a(r)$. 
For more complex models, additional geometric quantities are required to fully characterize the black hole interiors. 
For example, it was pointed out in Ref.~\cite{Arean:2024pzo} that when the Kasner singularity exists, $l\sqrt{K_{\mu\nu}K^{\mu\nu}}$ can provide more complete information, which is not significantly different from $|lK|$ in the case of spacelike singularities. 

\section{Complexity equals anything for multi-horizon black holes}\label{sec6}
\begin{figure}[htbp]
    \centering
    \begin{subfigure}[b]{0.45\textwidth}
        \centering
        \includegraphics[scale=0.15]{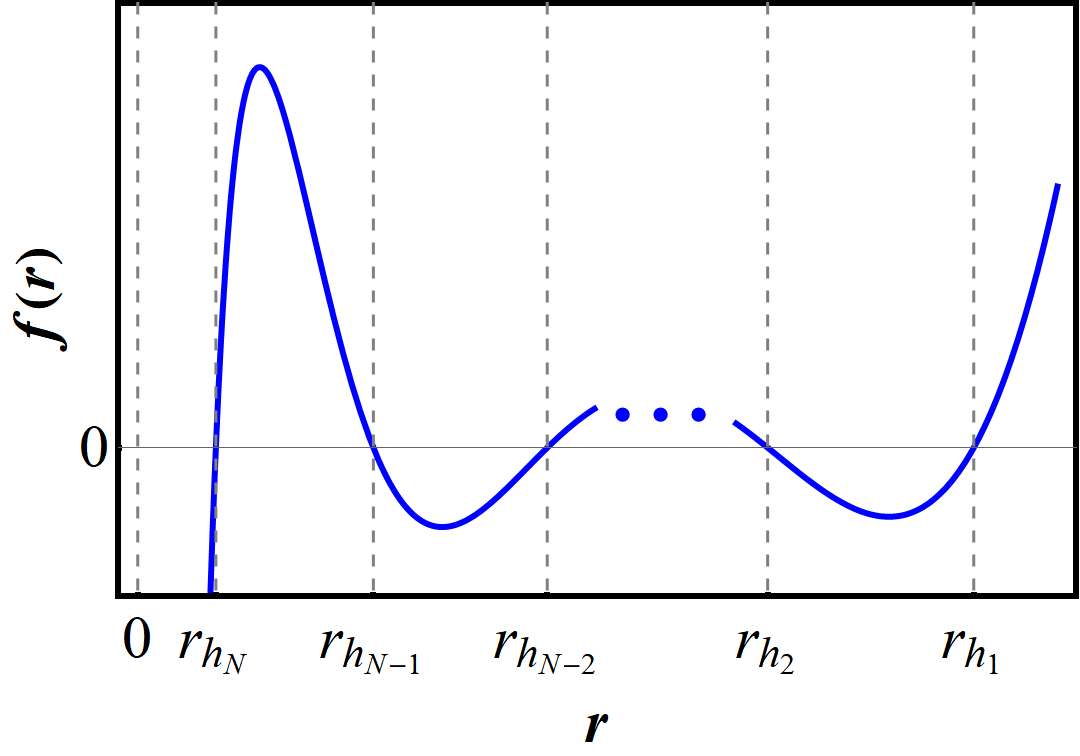}
        \caption[swx]{$N=2n-1$}
        \label{fr2n-1}
    \end{subfigure}
    \hspace{0.03\textwidth}
    \begin{subfigure}[b]{0.45\textwidth}
        \centering
        \includegraphics[scale=0.15]{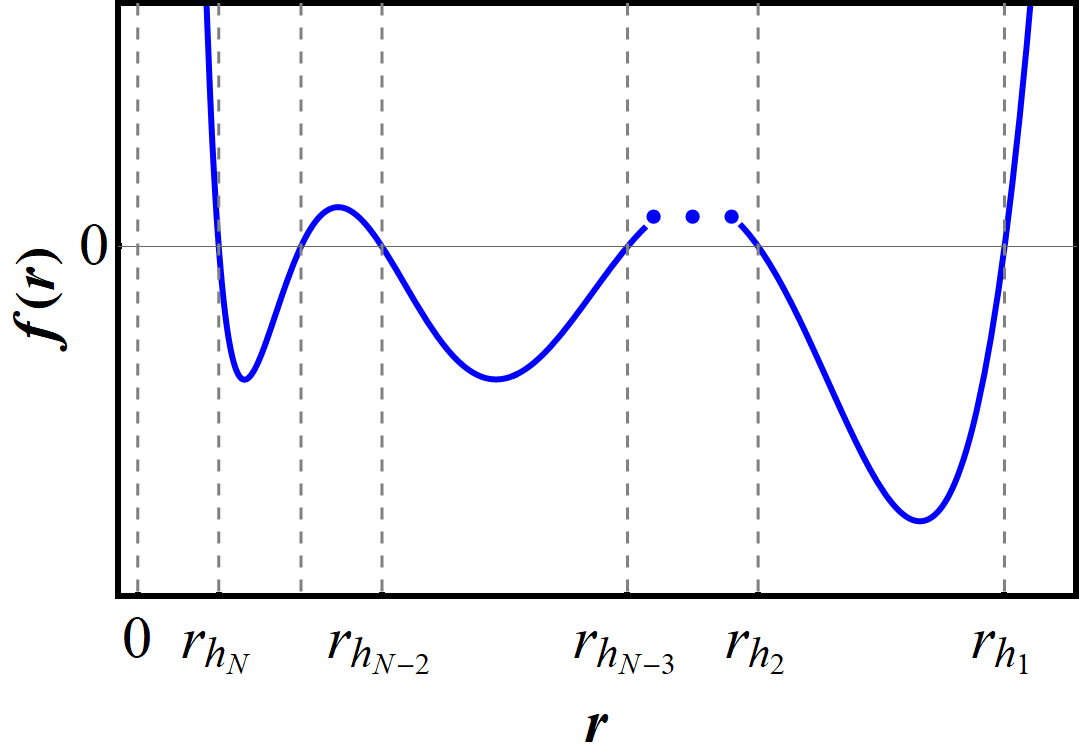}
        \caption[swx]{$N=2n$}
        \label{fr2n}
    \end{subfigure}
    \caption[swx]{An example of the AdS black hole with N horizons. 
    (a) An AdS black hole with an odd number of horizons has a spacelike curvature singularity at $r=0$.
    (b) An AdS black hole with an even number of horizons has a timelike curvature singularity at $r=0$.
    }
    \label{frN}
\end{figure}

\begin{figure}[htbp]
    \centering
    \begin{subfigure}[b]{0.45\textwidth}
        \centering
        \includegraphics[scale=0.15]{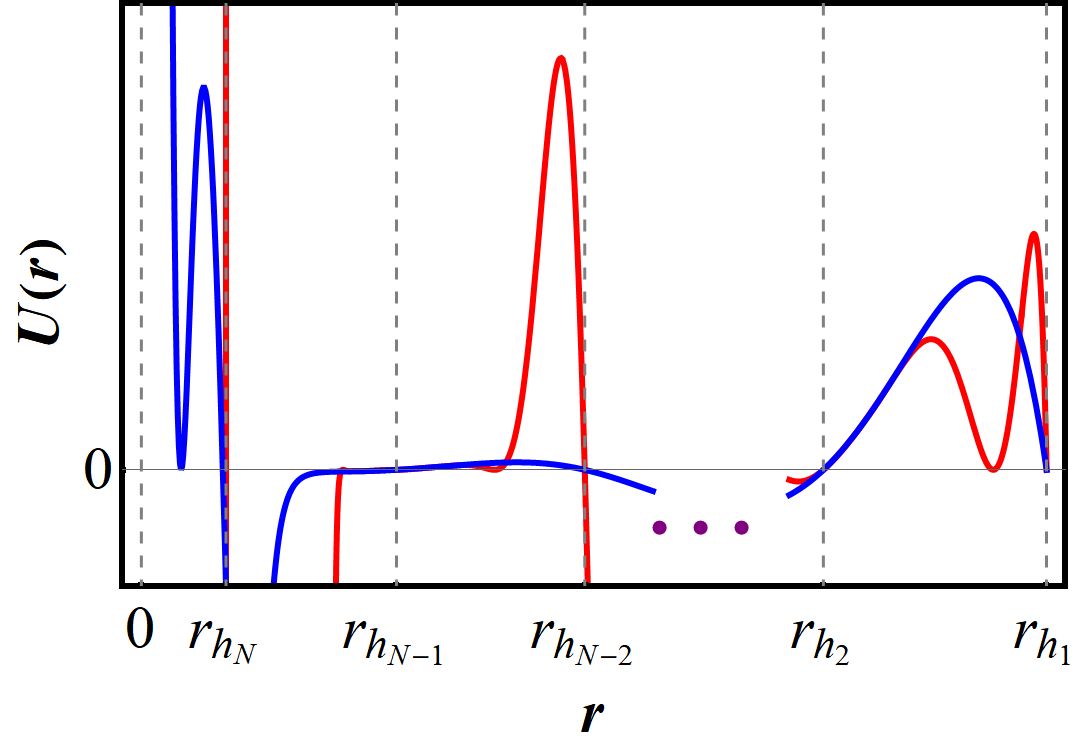}
        \caption[swx]{$N=2n-1$}
        \label{Ur2n-1}
    \end{subfigure}
    \hspace{0.05\textwidth}
    \begin{subfigure}[b]{0.45\textwidth}
        \centering
        \includegraphics[scale=0.15]{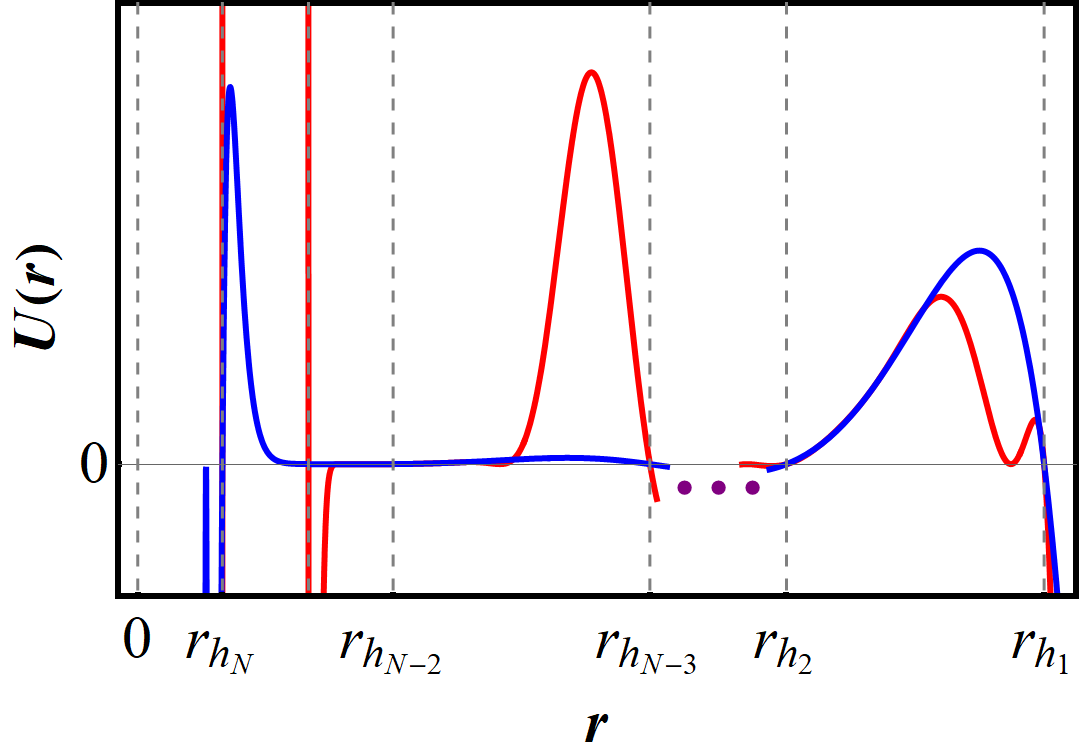}
        \caption[swx]{$N=2n$}
        \label{Ur2n}
    \end{subfigure}
    \caption[swx]{
        The effective potentials for multi-horizon black holes. 
    (a) An AdS black hole with an odd number of horizons. 
    (b) An AdS black hole with an even number of horizons. }
    \label{UrN}
\end{figure}

In this section, we consider a series of $(d+1)-$dimensional AdS black holes with multiple horizons, whose metric can be written as 
\begin{align}
    ds^{2}&=-f(r)dt^{2}+\frac{1}{f(r)}dr^{2}+r^{2}{d\Omega^{2}_{d-1}},\\
    f(r) &= g(r,r_{i})\,\frac{r^{2}}{L^{2}}\, \prod_{i=1}^{N} \left(1 - \frac{r_{i}}{r}\right),
\end{align}
where $g(r,r_{i})$ is a dimensionless function with $g(r,r_{i})|_{r\to\infty}\to1$, 
$N$ is the number of horizons, $r_{i}$ represents the the $i-$th horizon counting from the event horizon. 
In this case, we classify the discussion according to the characteristics of the curvature singularity, i.e., 
\begin{equation} 
    \left\{
    \begin{aligned}
        N=&2n-1\,\,\,:\,\,\,\text{Spacelike singularity}\\
        N=&2n\,\,\,:\,\,\,\text{Timelike singularity}\\
    \end{aligned},\,\,\, n=1,2,3,\cdots.
    \right.
\end{equation}
The most significant difference between these two cases is the asymptotic behavior of the blackening factor near the curvature singularity, i.e., 
\begin{equation}
    \lim_{r\to 0}f(r)\Big{|}_{N=2n-1}\to -\infty,\,\,\,\,\,\lim_{r\to 0}f(r)\Big{|}_{N=2n}\to +\infty.
\end{equation}
An intuitive example is given in Fig.~\ref{frN}. 
Considering the holographic complexity, we adopt the scalar function proposed in Ref.~\cite{Jiang:2023jti}, i.e., 
\begin{equation}
    a(r)=1+\sum_{j=1}^{K}(-1)^{j}\lambda_{j}(l^{4}C^{2})^{j}, 
\end{equation}
where $C^{2}$ is still the square of the Weyl tensor, $\lambda_{j}$ are dimensionless parameters, and $j=1,2, \cdots, K$. 
This construction allows us to probe different regions of the black hole interior by adjusting the number of the independent parameters $\lambda_{j}$. 
This is the advantage brought by the flexibility of the ``complexity equals anything'' proposal. 
This study focuses exclusively on the anomalous spacetime region of the black hole interior, that is, the part where $f(r)<0$. 
For the part where $f(r)>0$, the effective potential is always negative, which makes these regions impossible to be probed by holographic complexity. 
Figure \ref{UrN} shows two common examples of the effective potentials for the AdS black holes with $N$ horizons. 
We find that by adjusting the dimensionless parameters, we can always ensure the local maximum of the effective potential appear in any region where $f(r)<0$. 
On the other hand, changing $a(r)$ could construct a more complex effective potential, leading to a first-order phase transition in the complexity evolution process. However, this is not the focus of this paper. 
In addition, if we consider the codimension-zero observables, when $|K_{\Sigma_{+}}|\to\infty$, 
the future CMC slices will approach the Cauchy horizons, i.e., $r_{2k},\,\,\,k=1,2,3\cdots $. 
In other words, when
\begin{equation}
    \lim_{r\to r_{2k}^{+}} f(r)\to 0^{-},\,\,\,\,f'(r_{2k})<0, 
\end{equation}
$r_{f} \to r_{2k}^{+}$. 
Of course, if there is a spacelike curvature singularity at $r=0$, no matter how many horizons there are, the innermost CMC slice can always hug this singularity at late times. 

\section{Conclusion and discussion}\label{sec7}

In this paper, we have investigated the growth of holographic complexity in multi-horizon black holes using the ``complexity equals anything” proposal, with particular focus on the case involving three horizons. 
We discussed possible choices of the spacetime boundaries, 
and performed a detailed analysis of holographic complexity for the Bardeen-AdS class black hole. 
Through a detailed study of the extremal surfaces and their behavior near the singularity or Cauchy horizon, we obtained the late-time complexity growth. 

 {
In the three-horizon case, we demonstrated the superiority of the ``complexitiy equals anything" proposal in probing the black hole interiors, by analyzing both the evolution of codimension-one extremal surfaces and codimension-zero extremal regions, together with the dependence of late-time growth rates on the model parameters. }
Specifically, by varying the dimensionless parameter $\lambda_{2}$, we controlled how closely the extremal surface approaches the singularity and analyzed the asymptotic behavior of the complexity growth rate at late times as the extremal surface nears the singularity. 
On the other hand, by regulating the extrinsic curvature, we forced the future CMC slices to approach either the singularity or the Cauchy horizon. 
During this process, the high sensitivity of complexity to the choice of the scalar function $a(r)$ allowed us to distinguish different regions of the black hole interior through their characteristic complexity growth rates. 
Furthermore, we generalized these finding to black holes with more than three horizons. 
Due to the unique flexibility, the ``complexity equals anything'' proposal can always probe the spacetime regions where $f(r)<0$. 

In conclusion, this study establishes a framework for exploring the black hole interior with multiple horizons, 
making it possible to use the flexibility of the generalized complexity to explore more complete spacetime properties of the black hole interior. 
Future research could extend this methodology to more scenarios, such as black holes with multiple singularities or loop quantum black holes, 
and further investigate the role of various geometric quantities in characterizing interior spacetime structures.

Moreover, codimension-two observables are also important in the study of the black hole interior. 
The well-known codimension-two observables are Ryu-Takayanagi (RT) surface and Hubeny-Rangamani-Takayanagi (HRT) surface~\cite{Rangamani:2016dms}. 
For example, the entanglement entropy between the left and right CFTs in a TFD state is considered to be dual to the area of a time-dependent codimension-two extremal surface, which is called the Hartman-Maldacena (HM) surface~\cite{Hartman:2013qma}. 
This is a special case of the HRT surface~\cite{Lewkowycz:2013nqa,Dong:2016hjy}. 
It was noted in Ref.~\cite{Hartman:2013qma} that the entanglement entropy grows linearly after a time $t\sim \beta$ with $\beta$ being the inverse Hawking temperature, but it quickly reaches saturation. 
From a holographic point of view, the HM surface extends to the black hole interior until a critical time. 
After this, the holographic duality of the entangled entropy is replaced by two independent RT surfaces, which cannot enter the black hole interior.  
In other words, the codimension-two extremal surfaces can only cover a very limited region of the black hole interior. 
This is what Susskind mentioned in Ref.~\cite{Susskind:2014moa}: ``entanglement is not enough". 
When we consider a multi-horizon black hole, it is hard to extend the HM surface beyond the inner horizon. 
To enable the HM prescription to distinguish different left boundaries, 
we would need to modify the integrand of the codimension-two surface, which is similar to the ``complexity equals anything'' proposal~\cite{Belin:2022xmt}. 
This might be an interesting problem.

\section*{Acknowledgments}
We would like to thank Juan F. Pedraza, Le-Chen Qu and Shan-Ming Ruan for their very useful discussions and comments. 
We also thank Jing-Cheng Chang, Yi-Yang Guo, Shan-Ping Wu and Meng-Wei Xie for their selfless help. 
This work was supported by 
the National Natural Science Foundation of China (Grants No. 12475056, No. 12575055, No. 12275238 and No. 12247101), 
the Fundamental Research Funds for the Central Universities (Grant No. lzujbky-2025-jdzx07), 
the Natural Science Foundation of Gansu Province (No. 22JR5RA389, No.25JRRA799),  
and the `111 Center' under Grant No. B20063, Gansu Province's Top Leading Talent Support Plan. 
HYJ acknowledges the financial support provided by the scholarship granted by the Chinese Scholarship Council (CSC). 

\appendix
\section{Appendix: Generalized volume complexity constructed by matter content}\label{appa}
\begin{figure}[htbp]
    \centering
    \begin{subfigure}[b]{0.45\textwidth}
        \centering
        \includegraphics[scale=0.15]{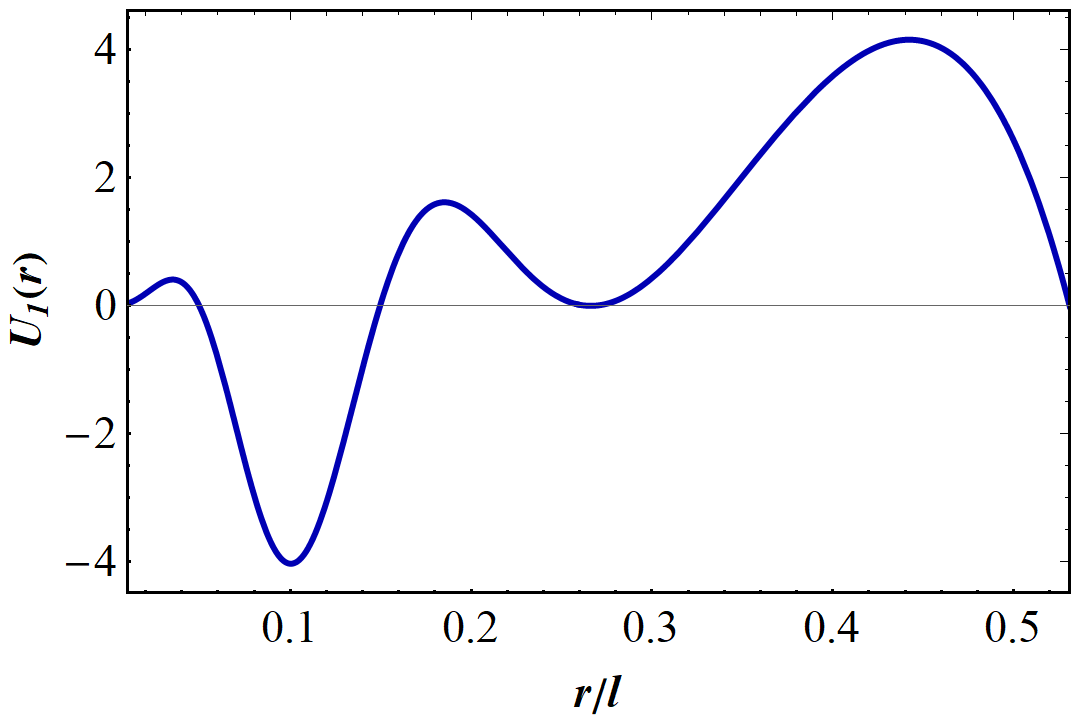}
        \caption[swx]{$U_{1}(r)=-f(r)a_{1}(r)^{2}r^{4}$.}
    \end{subfigure}
    \hspace{0.03\textwidth}
    \begin{subfigure}[b]{0.45\textwidth}
        \centering
        \includegraphics[scale=0.15]{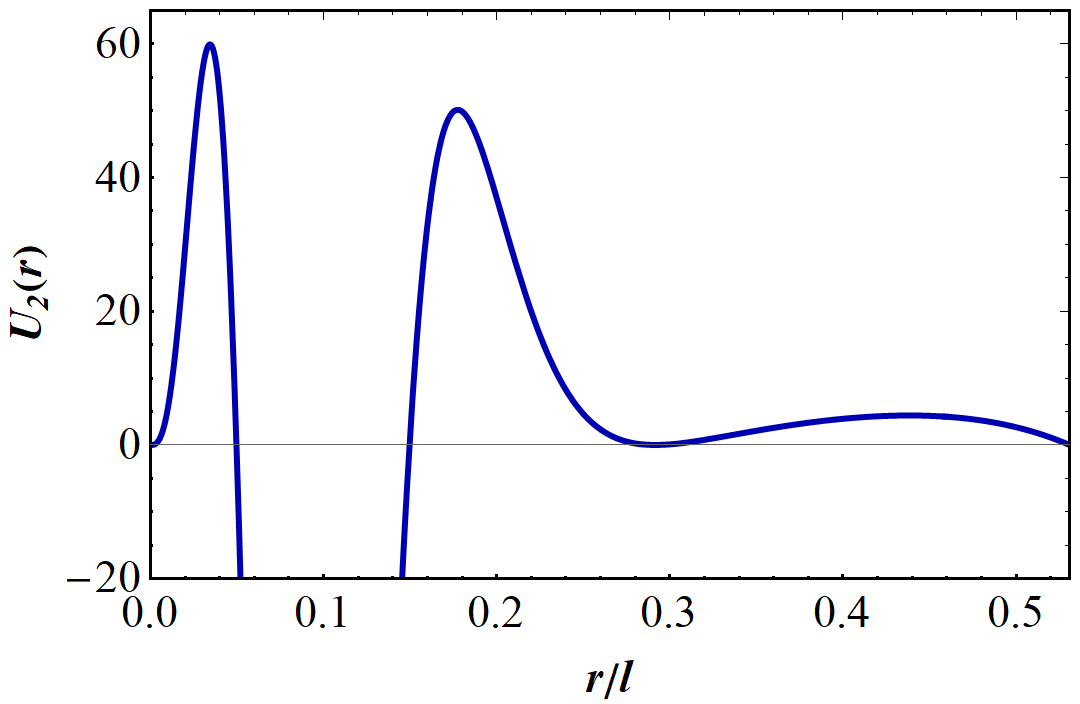}
        \caption[swx]{$U_{2}(r)=-f(r)a_{2}(r)^{2}r^{4}$.}
    \end{subfigure}
    \caption[swx]{
       The effective potentials with $ l = 5,~\alpha = 1,~s = 3,~M = 1.05~q^{3}/\alpha,~q = 1.3,~\lambda_{1}=0.1, \lambda_{2}=0.3$, where the Bardeen-AdS class black hole has three horizons. 
    }
    \label{Aur}
\end{figure}

Nonlinear electrodynamics is the key factor for the emergence of multi-horizon black holes, which has been discussed in detail in Ref.~\cite{Gao:2021kvr}. 
Then, a natural question is whether the gravitational observables constructed from the matter sectors can influence the geometric properties of holographic complexity discussed in the main text. 
To address this, we consider the following matter-built weights $a(r)$ to be inserted in the generalized-volume functional:
\begin{align}
    a_{1}(r)&=1+\lambda_{1}~\alpha~\mathcal{L}(\mathcal{F},\alpha)-\lambda_{2}~\alpha^{2}~\mathcal{L}(\mathcal{F},\alpha)^{2},\\
    a_{2}(r)&=1+\lambda_{1}~\alpha~\mathcal{L}(\mathcal{F},\alpha)-\lambda_{2}~\alpha^{3}~\mathcal{L}(\mathcal{F},\alpha)^{3}, 
\end{align}
where the Lagrangian density $\mathcal{L}(\mathcal{F},\alpha)$ can be reduced to 
\begin{equation}
    \mathcal{L}(\alpha)=\frac{12}{\alpha}\frac{q^{5}}{(q^{2}+r^{2})^{5/2}}.
\end{equation}
The corresponding effective potentials $U_{1,2}(r)=-f(r)a_{1,2}(r)^{2}r^{4}$ exhibit additional local maxima inside the Cauchy horizon for suitable parameters, 
allowing the extremal surface to extend into deeper interior regions, as shown in Fig,~\ref{Aur}. 
However, for the Bardeen-AdS class black hole, $\mathcal{L}(\mathcal{F},\alpha)\to \text{Const.}$ as $r \to 0$, 
so the matter term cannot drive the leftmost peak of $U(r)$ arbitrarily close to the singularity. 
In other words, these constructions allow the extremal surfaces to extend into the Reigon III, but they cannot be tuned to approach $r=0$. 
This is an expected conclusion: the matter term affects the structure of black hole interiors , but cannot directly capture the geometric characteristics of spacetime. 
Therefore, while matter-built constructions permit penetration beyond the inner horizon, curvature-built choices remain more effective as singularity probes. 

\end{document}